\documentclass[12pt,a4paper]{article}

\usepackage[utf8]{inputenc}

\usepackage{jheppub}

\usepackage{amsmath}
\usepackage{amsfonts}
\usepackage{amssymb}

\usepackage[retainorgcmds]{IEEEtrantools}

\usepackage{multirow}
\usepackage{multicol}
\usepackage{graphicx}
\usepackage{tikz}
\usepackage{niceframe}
\usepackage{float}

\usepackage{hyperref}

\usepackage{multirow}

\usepackage[mathscr]{eucal}
\usepackage{enumerate}
\numberwithin{equation}{section}

\def\a{{\alpha}}      
\def\b{{\beta}}
\def\g{{\gamma}}
\def\d{{\delta}}
\def\r{{\rho}}
\def\s{{\sigma}}

\def\e{{\epsilon}}

\def\th{{\theta}}
\def\S{{\Sigma}}
\def\L{{\Lambda}}

\def\ad{{\dot{\alpha}}}  
\def\bd{{\dot{\beta}}}
\def\gd{{\dot{\gamma}}}

\def\ed{{\bar{\epsilon}}}
\def\thd{{\bar{\theta}}}
\def\Sd{{\bar{\Sigma}}}
\def\Ld{\bar{\Lambda}}

\def\N{{{\bf {\cal N}}}}

\def\n{{\eta}}

\def\D{{\rm D}}         
\def\Dd{{\bar{\rm D}}}
\def\pa{\partial}

\def\N{{\bar{N}}}

\def\mS{{\mathscr{S}}}

\def\N{{\mathscr{N}}}

\def\[{\left[}
\def\]{\right]}

\def\be{\begin{equation}}
\def\ee{\end{equation}}
\def\bea{\begin{IEEEeqnarray*}}
\def\eea{\end{IEEEeqnarray*}}

\def\n{\IEEEyesnumber}            
\def\sn{\IEEEyessubnumber}

\title{Complex linear superfields, Supercurrents and Supergravities}
\author{P.~Ko\v{c}\'{i}}
\emailAdd{pavelkoci@mail.muni.cz}

\author{, K.~Koutrolikos}
\emailAdd{kkoutrolikos@physics.muni.cz}

\author{\! and \!}

\author{R.~von~Unge}
\emailAdd{unge@physics.muni.cz}

\affiliation{Institute for Theoretical Physics and Astrophysics\\
Masaryk University,\\ 611 37 Brno, Czech Republic}

\abstract{
We present expressions for the supercurrents generated by a generic $4D,~\mathcal{N}=1$ theory of complex linear superfield $\S$. We verify that these expressions satisfy the appropriate superspace conservation equations. Furthermore, we discuss the component projection in order to derive expressions for the energy-momentum tensor, the supersymmetry current and the R-symmetry current when available. In addition, we discuss aspects of the coupling of the theory to supergravity. Specifically, we present a straightforward method to select the appropriate formulations of supergravity that one must use in order to do the coupling. This procedure is controlled by a superfield X originating from the Super-Poincar\'{e} invariance of the theory. We apply these results to examples of theories with higher derivative terms. 
}

\begin{document}

\maketitle



\section{Introduction}
~~~~An interesting feature of Superspace Supersymmetric field theory is the existence of alternative representations\hyphenation{re-pre-se-nta-tions} of various well-known supermultiplets. These \emph{variant}~\cite{vsr} descriptions, although describing the same on-shell degrees of freedom, provide different auxiliary field structures. Examples of variant representations are the known different (minimal\hyphenation{mi-ni-mal} and non-minimal) formulations of $4D,~\mathcal{N}=1$ supergravity where the variants appear in terms of alternative compensating superfields.

The archetypical example of this phenomenon is the scalar supermultiplet. The economical and most frequent superspace description is via a chiral superfield, but the same on-shell degrees of freedom can also be described in terms of a complex linear superfield \cite{Siegel cld, Lindstrom cld, Deo cld}. Moreover, these two descriptions are connected by a superspace duality procedure which gives a concrete prescription for how they are related\footnote{So far it was believed that variant representations and especially the linear-chiral duality were a feature of the low spin theories. However recently \cite{HS cld} a higher-spin generalization was demonstrated in $3D$.}. Through this duality, a wide class of theories can be described in superspace using either chiral superfields or alternatively complex linear superfields, hence one might be tempted to draw the conclusion that this is true for any theory. It has recently become clear that this is not the case since in certain theories with higher derivative terms some of the auxiliary components can become propagating and thus change the degrees of freedom of the theory. This fact motivated the use of complex linear superfield as a prime candidate for various phenomenological models regarding spontaneous supersymmetry breaking \cite{susybr1, susybr2, susybr5, susybr3, susybr4, susybr6} that is, so far, much less understood than the standard mechanisms. For that reason it would be desirable to study aspects of theories of complex linear superfields.

In this paper we focus on the supercurrents that can be generated by these theories and the coupling to supergravity. In a supersymmetric theory, the currents themselves form a supermultiplet which can be encoded in a superfield and this has been used extensively \cite{sc1, sc2, sc3}. To determine the supercurrent multiplet of a generic theory we could follow the superfield Noether procedure developed in \cite{sNp1, sNp2}. Alternatively, we know that if a superfield description of the coupling of the theory in question to supergravity is available, then the supercurrent multiplet can be calculated from the equations of motion of the supergravity superfields in the limit where they vanish \cite{ideas, GGRS}. Of course, in the linearised limit which defines the supercurrent multiplet, the two methods match since the Noether procedure gives the coupling to supergravity. An obvious remark is that the supercurrent multiplet must match the superfields required in a specific formulation of supergravity. So we immediately know how many superfields and of what type (vectors, spinors, scalars, real or not) we should expect to participate in the description of the supercurrent multiplet.

For our case, this translates to consider the change of $\S$ under  linearized superdiffeomrphism and perform one Noether iteration. The results are that for an arbitrary $4D,~\mathcal{N}=1$ theory of $\S$ we:\\
1. Identify a set of objects $\{\N_{\a\ad},\N_{\ad},\N,\mathscr{M}\}$
which depending on the formulation of supergravity (minimal, non-minimal) we use, they generate the appropriate superfields that will describe the corresponding supercurrent multiplet. For each one of them we give an explicit $\S$ dependence.\\
2. Verify that the expressions for the supercurrent multiplets generated by the above process satisfy the relevant superspace conservation equations.\\
3. Provide expressions for energy-momentum tensor, supersymmetry current and R-symmetry current (if present) by projecting the superspace conservation equations to spacetime.\\
4. Propose a method to determine the formulations of supergravity which are compatible with a given theory and therefore could be used for coupling. This method is controlled by a superfield X which comes from the structure of the supercurrent multiplet under rigid Super-Poincar\'{e} transformations

The paper is organized as follows. In section 2, we start by considering rigid Super-Poincar\'{e} transformations. As expected the Super-Poincar\'{e} invariance enforces a specific structure on the corresponding supercurrent multiplets which will be used extensively in the following sections. In sections 3, we discuss the action of the superdiffeomorphism group on $\S$. Namely, the compatibility of the trasnformation with the linearity constraint will fix the structure of most of the transformation parameters. We will discover that the most general transformation allowed is not constrained enough and the demand to couple the theory to pure supergravity will fix the rest of the freedom. In section 4, we initiate the Noether procedure for the above transformation and discover a list of potential supercurrents $\{\N_{\a\ad},\N_{\ad},\N,\mathscr{M}\}$. This is a by-product of the fact that there are more than one formulations of supergravity and we haven't made a choice yet. In section 5, we remind the reader of the different possible supergravity formulations and finish the Noether procedure. For each choice of supergravity the Noether procedure will lead to a set of constraints that need to be imposed. The results are
1. the supercurrent multiplets and 2. the necessary $X$ constraint that is required in order to be possible the coupling of the theory with the specific version of supergravity. In section 6, we confirm that the supercurrent multiplets of section 5 conform with the appropriate superspace conservation equation\cite{sc1, sc2, sc3}, in accordance with the supergravity choice. We do that by deriving the corresponding Bianchi identities and take their on-shell limit.
Furthermore, we demonstrate that for the case of new-minimal supergravity, the necessary conditions we derived in section 5 correspond to having R-symmetry invariance in the theory. This explains the fact that the new-minimal superspace conservation equation gives the spacetime conservation equation of the R-symmetry current. Finally, we project to components and derive the corresponding set of spacetime conservation equations which involve the energy-momentum tensor, the supersymmetry current and the R-symmetry current for the case of new-minimal. In section 7, we apply all the above results to three specific examples of theories: (i) the (almost) free theory, (ii) an interacting theory with higher derivatives and (iii) an interacting theory with higher derivatives that has supersymmetry breaking solutions. Section 8 has the concluding comments.

%
%
\section{Rigid Super-Poincar\'{e} Noether Procedure}
~~~~All superspace formulated theories make manifest their invariance under rigid Super-Poincar\'{e} transformations. However, this restricts the structure of the supercurrents of the theory. To see this we perform a Super-Poincar\'{e} transformation parametrized by $a_{\a\ad}$, the symmetric $\omega_{\a\b},~ \omega_{\ad\bd}$ and $\e_{\a},~ \ed_{\ad}$\footnote{We use {\it{Superspace}} \cite{GGRS} conventions.}
\bea{l}\n
x'^{\a\ad}=x^{\a\ad}+a^{\a\ad}-\tfrac{i}{2}\e^{\a}\thd^{\ad}-\tfrac{i}{2}\ed^{\ad}\thd^{\a}+\tfrac{1}{2}\omega^{\a\b}x_{\b}{}^{\ad}+\tfrac{1}{2}x^{\a}{}_{\bd}\omega^{\bd\ad} ~, \\
\th'^{\a}=\th^{\a}+\tfrac{1}{2}\omega^{\a\b}\th_{\b}+\e^{\a} \, , \\
\thd'^{\ad}=\thd^{\ad}+\tfrac{1}{2}\omega^{\ad\bd}\thd_{\bd}+\ed^{\ad}~.
\eea
The transformation of $\S$ is:
\bea{l}\n
\d_{S.P.}\S=\Delta_{S.P.}^{\a}\D_{\a}\S+\Delta_{S.P.}^{\ad}\Dd_{\ad}\S+i\Delta_{S.P.}^{\a\ad}\pa_{\a\ad}\S~, \\
\Delta_{S.P.}^{\a}=-(\e^{\a}+\tfrac{1}{2}\omega^{\a\b}\th_{\b}) \, ,\\
\Delta_{S.P.}^{\ad}=-(\ed^{\ad}+\tfrac{1}{2}\omega^{\ad\bd}\thd_{\bd})\, ,\\
\Delta_{S.P.}^{\a\ad}=ia^{\a\ad}+\e^{\a}\thd^{\ad}+\ed^{\ad}\th^{\a}+\tfrac{i}{2}\omega^{\a\b}\{x_{\b}{}^{\ad}-\tfrac{i}{2}\th_{\b}\thd^{\ad}\}+\tfrac{i}{2}\{x^{\a}{}_{\bd}-\tfrac{i}{2}\thd_{\bd}\th^{\a}\}\omega^{\bd\ad}~.
\eea
{}From this follows
\bea{ll}\n\label{SPconst}
\Dd_{(\bd}\Delta^{S.P.}_{\a\ad)}=0~,~~&~~\bar{\Delta}^{S.P.}_{\a\ad}=-\Delta^{S.P.}_{\a\ad} ~,\\
\Delta^{S.P.}_{\a}=-\tfrac{1}{2}\Dd^{\ad}\Delta^{S.P.}_{\a\ad}~,~~&~~\Delta^{S.P.}_{\ad}=\tfrac{1}{2}\D^{\a}\bar{\Delta}^{S.P.}_{\a\ad}~,\\
\D^{\a}\Delta^{S.P.}_{\a}=0~,~~&~~\Dd^{\ad}\Delta^{S.P.}_{\ad}=0~,
\eea
which are compatible with the linearity constraint of $\S$. Notice that
$\Delta^{S.P.}_{\a}$, $\Delta^{S.P.}_{\ad}$ are not independent and can be derived from $\Delta^{S.P.}_{\a\ad}$.
The variation of a general action $S_o[\S,\Sd]$ under the global S.P. transformation \hyphenation{tra-nsfo-rma-tion} is:
\bea{rl}
\delta_{S.P.} S_o[\S,\Sd]=\int d^8z~~&\Delta^{\a\ad}_{S.P.}\left\{J_{\a\ad}-\bar{J}_{\a\ad}\right\}\\
~~~=\int d^8z~~&ia^{\a\ad}\left\{J_{\a\ad}-\bar{J}_{\a\ad}\right\}\n\label{dSSP}\\
+~&\e^{\a}\thd^{\ad}\left\{J_{\a\ad}-\bar{J}_{\a\ad}\right\}+c.c.\\
+~&\tfrac{i}{2}\omega^{\a\b}
(x_{\b}{}^{\ad}-\tfrac{i}{2}\th_{\b}\thd^{\ad})
\left\{J_{\a\ad}-\bar{J}_{\a\ad}\right\}+c.c.~,
\eea
where
\bea{l}
J_{\a\ad}=i\pa_{\a\ad}\S~T_o-\tfrac{1}{2}\Dd_{\ad}\left(\D_{\a}\S~T_o+\D_{\a}\Sd~\bar{T}_o\right)\n\label{J}~,
\eea
and $T_o=\tfrac{\d S_o}{\d\S}$ is the variation of the action with respect to $\S$.
Keep in mind that due to the linearity constraint of $\S$, its equation of motion is $\Dd_{\ad}T_o=0$.
For $S_o[\S,\Sd]$ to be Super-Poincar\'{e} invariant\hyphenation{inva-ri-ant}, the combinations $J_{\a\ad}-\bar{J}_{\a\ad}$, $\thd^{\ad}\left\{J_{\a\ad}-\bar{J}_{\a\ad}\right\}$ and $\left(x_{(\b}{}^{\ad}-\tfrac{i}{2}\th_{(\b}\thd^{\ad}\right)\left\{J_{\a)\ad}-\bar{J}_{\a)\ad}\right\}$ must be total superspace derivatives. In other words, there must exist superfields $A_{\b\a\ad}, B_{\b\a}, C_{\a\ad}, F_{\g\b\a}, G_{\b\a\ad}$ such that:
\begin{enumerate}
\item $a_{\a\ad}$ term:~~$J_{\a\ad}-\bar{J}_{\a\ad}=\D^{\b}A_{\b\a\ad}-\Dd^{\bd}\bar{A}_{\a\bd\ad}~,$
\item $\e^{\a}$ term: ~~$\thd^{\ad}\left\{J_{\a\ad}-\bar{J}_{\a\ad}\right\}=\D^{\b}B_{\b\a}+\Dd^{\bd}C_{\a\bd}~,$
\item $\omega^{\a\b}$ term:~~$\left(x_{(\b}{}^{\ad}-\tfrac{i}{2}\th_{(\b}\thd^{\ad}\right)\left\{J_{\a)\ad}-\bar{J}_{\a)\ad}\right\}=\D^{\g}F_{\g\b\a}+\Dd^{\gd}G_{\b\a\gd}~.$
\end{enumerate}
The result is that for any theory of complex linear superfields $S_o[\S,\Sd]$ the imaginary part of $J_{\a\ad}$ can always be written in the following form: 
\bea{l}\n\label{K}
\mathcal{K}_{\a\ad}\equiv J_{\a\ad}-\bar{J}_{\a\ad}=\D^{\b}\Omega_{\b\a\ad}-\Dd^{\bd}\bar{\Omega}_{\a\bd\ad}+\D_{\a}\Dd_{\ad}X+\Dd_{\ad}\D_{\a}\bar{X}~,
\eea
for some superfields $X$ and $\Omega_{\b\a\ad}$ with $\Omega_{\b\a\ad}=\Omega_{\a\b\ad}$.
However, due to (\ref{dSSP}) and $\Dd_{(\bd}\Delta^{S.P.}_{\a\ad)}=0$ it is obvious that $J_{\a\ad}$ is not uniquely defined and there is a redundancy. This freedom resolves to the identification
\bea{l}\n
\mathcal{K}_{\a\ad}\sim\mathcal{K}_{\a\ad}+\D^{\b}R_{(\b\a)\ad}-\Dd^{\bd}\bar{R}_{\a(\bd\ad)}~.
\eea
We can exploit this by choosing $R_{(\b\a)\ad}=-\Omega_{\b\a\ad}$ and simplify the expression for $\mathcal{K}_{\a\ad}$ to be\footnote{This was first shown in \cite{sNp1} and later in \cite{sNp2}.}
\bea{l}\n\label{Krd}
\mathcal{K}_{\a\ad}=\D_{\a}\Dd_{\ad}X+\Dd_{\ad}\D_{\a}\bar{X}~.
\eea
As we will see later, the superfield $X$ plays a key role in determining
the formulation of supergravity which must be used in order to couple the theory.
%
%
%
%
\section{Superdiffeomorphism group action}
~~~~Now we move on to the more interesting case of local super-diffeomorphisms. As we mentioned in the introduction, for the purpose of finding the supercurrents working to linear order is enough and we have to consider the linearised transformation of $\S$:
\be
\d\S=\Delta^\a \D_{\a}\S+\Delta^{\ad} \Dd_{\ad}\S 
+ i\Delta^{\a\ad} \pa_{\a\ad}\S+\Delta\S\label{dS}~.
\ee
Anticipating the fact that  
$\S$ may not be a scalar but a \emph{density}, we have introduced an additional term $\Delta\S$ giving a complex scaling of $\S$. After all, the super conformal group is naturally included in the super diffeomorphism group.
In the following sections, we will discover that $\Delta$ plays a very important role in the story of coupling the $\S$ theory to supergravity.

We need to make sure that the above transformation respects the linearity constraint of $\Sigma$. In other words, the set of parameters $\Delta$s must satisfy the constraints:
\bea{l}\n\label{const}
\Dd^2\d\S=0~\Rightarrow~\begin{cases}
\Delta_{\a}=-\tfrac{1}{2}\Dd^{\ad}\Delta_{\a\ad}~,\\
\Dd_{(\bd}\Delta_{\a\ad)}=0~,\\
\Dd^2\Delta_{\ad}+\Dd_{\ad}\Delta=0~.
\end{cases}
\eea
The most general solution of the above constraints is
\bea{l}\n\label{solconst}
\Delta_{\a\ad}=\Dd_{\ad}\L_{\a}~,\sn\label{sol1}\\
\Delta_{\a}=-\Dd^2\L_{\a}~,\sn\label{sol2}\\
\Delta=\Dd^{\ad}\Delta_{\ad}+\varphi~,~\Dd_{\bd}\varphi=0\sn\label{sol3}~.
\eea
We see that the parameters are given in terms of the two unconstrained spinorial superfields $\Lambda_\a,\Delta_\ad$ and the chiral field $\varphi$. If $\Delta=0$ then $\varphi$ is no longer independent and $\Delta_\ad$ is further constrained, $\Dd^2 \Delta_\ad = 0$. The conclusion is that the most general transformation\hyphenation{trans-for-ma-tion} of $\S$ allowed is:
\be
\d\S=-\Dd^2\L^{\a}\D_{\a}\S+\Delta^{\ad}\Dd_{\ad}\S+i\Dd^{\ad}\L^{\a}\pa_{\a\ad}\S+\left(\Dd^{\ad}\Delta_{\ad}+\varphi\right)\S\label{S-transf}~.
\ee

However, the demand for an invariant action that couples $\S$ with supergravity dictates that $\Delta_{\a\ad}, \Delta_{\a}, \Delta_{\ad}, \Delta$ and $\varphi$ must be functions of the superfield parameters that appear in the transformation of the supergravity superfields. Equations (\ref{sol1}) and (\ref{sol2}) completely fix this correspondence for $\Delta_{\a\ad}$ and $\Delta_{\a}$, but on the other hand (\ref{sol3}) gives a lot of flexibility regarding $\Delta,\Delta_{\ad}$ and $\varphi$. The most general ansatz we can do for $\Delta$ regarding its $\L_{\a}$ dependence is 
\be
\Delta=\kappa_1\D^{\a}\Dd^2\L_{\a}+\kappa^{*}_{2}\Dd^{\ad}\D^2\bar{\L}_{\ad}+\lambda_1\Dd^2\D^{\a}\L_{\a}+\lambda^{*}_2\D^2\Dd^{\ad}\bar{\L}_{\ad}~,
\ee
for some arbitrary parameters $\kappa_1,\kappa_2,\lambda_1,\lambda_2$.
Substituting the above to (\ref{sol3}) and taking into account the chiral property of $\varphi$ we get the following parametrization
\bea{l}\n\label{GenPar}
\Delta_\ad = (-1+c)\D^2\Ld_\ad - \kappa\D^\a\Dd_\ad\L_\a + \Dd^\bd\Ld_{\ad\bd} + \Dd_\ad\Ld ~,~\L_{\a\b}=\L_{\b\a},\sn\\
\Delta = (-1+c)\Dd^{\ad}\D^2\Ld_{\ad} +\kappa\D^\a\Dd^2\L_\a + \lambda\Dd^2\D^\a\L_\a~,\sn\\
\varphi = (\lambda-\kappa)\Dd^2\D^\a\L_\a - 2\Dd^2\Ld~,\sn
\eea
where we have also conveniently redefined the remaining parameters.

The conclusion of this section is that the construction of an invariant action that couples the matter theory of $\S$ with linearized supergravity must be based upon (\ref{S-transf}) together with (\ref{GenPar}).
%
%
%
%
\section{Prelude to an invariant theory}
~~~~We start with a generic action for $\S$
\be
S_o=\int d^8z~\mathcal{L}_{o}(\S,\Sd)~,
\ee
and calculate the change of it under the above transformation. We get
\bea{lll}
\delta S_o &=\int d^8z ~&~~\Delta^{\a\ad}\left\{i\pa_{\a\ad}\S ~T_o-\tfrac{1}{2}\Dd_{\ad}\left(\D_{\a}\S ~T_o\right)\right\}+c.c.\\
& &+\Delta^{\ad}\left\{\Dd_{\ad}\S ~T_o\right\}+c.c.\n\\
& &+\[\Dd^{\ad}\Delta_{\ad}+\varphi\]\S T_o+c.c.~.
\eea
Using (\ref{J}) this can be written as:
\bea{lll}
\delta S_o &=\int d^8z ~&~~\tfrac{1}{2}\left(\Delta^{\a\ad}+\bar{\Delta}^{\a\ad}\right)\left\{J_{\a\ad}+\bar{J}_{\a\ad}\right\}\\
& &+\tfrac{1}{2}\left(\Delta^{\a\ad}-\bar{\Delta}^{\a\ad}\right)\left\{J_{\a\ad}-\bar{J}_{\a\ad}\right\}\n\\
& &+\[\Delta^{\ad}+\tfrac{1}{2}\D_{\a}\bar{\Delta}^{\a\ad}\]\Dd_{\ad}\S ~T_o+c.c.\\
& &+\[\Dd^{\ad}\Delta_{\ad}+\varphi\]\S T_o+c.c.~.
\eea
Now we make use of the specific structure (\ref{Krd}) of $J_{\a\ad}-\bar{J}_{\a\ad}$\footnote{We must keep in mind that in order to get equation (\ref{Krd}) we have redefined $J_{\a\ad}$ to $J_{\a\ad}+\Dd^{\bd}\bar{\Omega}_{\a\bd\ad}$. This mean that the $\Omega_{\b\a\ad}$ dependence will disappear from $J_{\a\ad}-\bar{J}_{\a\ad}$ but it will appear in the $J_{\a\ad}+\bar{J}_{\a\ad}$. Alternatively, we can forget all about (\ref{Krd}) and use the full equation (\ref{K}) together with the constraint $\Dd_{(\bd}\Delta_{\a\ad)}=0$.}:
\bea{lll}
\delta S_o &=\int d^8z ~&-\tfrac{1}{2}\left(\D^{\a}\bar{\L}^{\ad}-\Dd^{\ad}\L^\a\right)\left\{T_{\a\ad}+\D^{\b}\Omega_{\b\a\ad}
+\Dd^{\bd}\bar{\Omega}_{\a\bd\ad}\right\}\\
& &+\D^{\a}\Dd^2\L_{\a}\left\{\tfrac{1}{2}X-\bar{X}\right\}+c.c.\\
& &+\Dd^2\D^{\a}\L_{\a}\left\{\tfrac{1}{2}X\right\}+c.c.\n\label{dSo}\\
& &+\[\Delta^{\ad}+\D^2\bar{\L}^{\ad}\]\Dd_{\ad}\S ~T_o+c.c.\\
& &+\[\Dd^{\ad}\Delta_{\ad}+\varphi\]\S T_o+c.c.~,
\eea
where $T_{\a\ad}=J_{\a\ad}+\bar{J}_{\a\ad}$. This expression for the deformation of the action seems to be clear and unambiguous. However, this is not true because the terms $\D^{\a}\Dd^2\L_{\a}$,$\Dd^2\D^{\a}\L_{\a}$ and $\D_{\a}\bar{\L}_{\ad}-\Dd_{\ad}\L_{\a}$ are not independent under the integration sign. We have the freedom to perform integrations by part and transform them among themselves. This can be demonstrated by the following identity:
\bea{c}\n\label{IT}
\int d^8z\left[\D^{\a}\Dd^2\L_{\a}\left\{W+\tfrac{1}{2}\bar{W}\right\}
+\Dd^2\D^{\a}\L_{\a}\left\{\tfrac{1}{2}\bar{W}\right\}\right]+c.c.~=\\
=~\int d^8z\left(\D^{\a}\bar{\L}^{\ad}-\Dd^{\ad}\L^{\a}\right)\left\{\tfrac{1}{2}\Dd_{\ad}\D_{\a}W-\tfrac{1}{2}\D_{\a}\Dd_{\ad}\bar{W}\right\}~,
\eea
for any superfield $W$. From the point of view of a Lagrangian description, this is the argument behind the existence of the improvement terms that can be used in order to change the structure of the supercurrent and the conservation equations. Different theories coupled to different supergravities will require different improvement terms. Therefore we add a general improvement term parametrized by the superfield $W$ and the variation of the matter system takes the form:
\bea{lll}
\hspace{-1ex}\d S_o &=\hspace{-1ex}\int\hspace{-1ex}d^8z &\hspace{0.5ex}-\tfrac{1}{2}\left(\D^{\a}\bar{\L}^{\ad}-\Dd^{\ad}\L^\a\right)\left\{T_{\a\ad}+\D^{\b}\Omega_{\b\a\ad}
+\Dd^{\bd}\bar{\Omega}_{\a\bd\ad}+\Dd_{\ad}\D_{\a}W
-\D_{\a}\Dd_{\ad}\bar{W}\right\}\\
& &+\D^{\a}\Dd^2\L_{\a}\left\{\tfrac{1}{2}X-\bar{X}+W+\tfrac{1}{2}\bar{W}\right\}+c.c.\\
& &+\Dd^2\D^{\a}\L_{\a}\left\{\tfrac{1}{2}X+\tfrac{1}{2}\bar{W}\right\}+c.c.\n\label{dSopf}\\
& &+\[\Delta^{\ad}+\D^2\bar{\L}^{\ad}\]\Dd_{\ad}\S~T_o+c.c.\\
& &+\[\Dd^{\ad}\Delta_{\ad}+\varphi\]\S T_o+c.c.~.
\eea
Finally, using (\ref{GenPar}) we may write
\bea{lll}
\hspace{-1ex}\d S_o &=\hspace{-1ex}\int\hspace{-1ex}d^8z ~&-\tfrac{1}{2}\left(\D^{\a}\bar{\L}^{\ad}-\Dd^{\ad}\L^\a\right)~\N_{\a\ad}\\
& &+\D^{\a}\Dd^2\L_{\a}~\N+c.c.\\
& &+\Dd^2\D^{\a}\L_{\a}~\mathscr{M}+c.c.\n\label{dSof}\\
& &+\left[c\D^2\bar{\L}^{\ad}-\kappa\D^{\a}\Dd^{\ad}\L_{\a}-\Dd_{\bd}\bar{\L}^{\bd\ad}+\Dd^{\ad}\bar{\L}\right]~\N_{\ad}+c.c.~,
\eea
with the following definitions:
\bea{l}\n
\N_{\a\ad}\equiv~T_{\a\ad}+\D^{\b}\Omega_{\b\a\ad}
+\Dd^{\bd}\bar{\Omega}_{\a\bd\ad}+\Dd_{\ad}\D_{\a}W
-\D_{\a}\Dd_{\ad}\bar{W}~,\sn\\
\N_{\ad}\equiv~\Dd_{\ad}\S~T_o~,\sn\label{Nad}\\
\N\equiv~\tfrac{1}{2}X-\bar{X}+W+\tfrac{1}{2}\bar{W}+\kappa\S T_o+(-1+c^\star)\Sd\bar{T}_o~,\sn\\
\mathscr{M}\equiv~\tfrac{1}{2}X+\tfrac{1}{2}\bar{W}+\lambda\S T_o~.\sn
\eea
Expression (\ref{dSof}) includes all the information required. The parameters $\kappa,c,\lambda$, the superfield $W$ and the combinations of $\N_{\a\ad},\N_{\ad},\N,\mathscr{M}$ that will eventually give the supercurrents must be determined in accordance with the choice of the supergravity formulation we want to couple to our theory.
%
%
%
%
\section{Supercurrents and Supergravities}
~~~~Given equation (\ref{dSof}), we want to find interaction terms $S_{int}[\S,\Sd,H_{\a\ad},C]$ such that $S_o[\S,\Sd] + S_{int}[\S,\Sd,H_{\a\ad},C]$ will be invariant to linear order.
At this point, it will be useful to review the various options that we have for irreducible supergravity theories (see \cite{ideas,GGRS} for reviews). 
\begin{enumerate}
\item \underline{Old-minimal}\cite{om1,om2,om3}: $\d H_{\a\ad}=\D_{\a}\bar{L}_{\ad}-\Dd_{\ad}L_{\a}~,~$
$\d\s=\Dd^2\D^{\a}L_{\a}~,\s$ is chiral.
\item \underline{New-minimal}\cite{newm1,newm2,newm3}: $\d H_{\a\ad}=\D_{\a}\bar{L}_{\ad}-\Dd_{\ad}L_{\a}~,~$
$\d U=\D^{\a}\Dd^2L_{\a}+\Dd^{\ad}\D^2\bar{L}_{\ad}~,U$ is real, linear. $U=\D^{\a}\psi_{\a}+\Dd^{\ad}\bar{\psi}_{\ad}~,~\Dd_\ad\psi_\a=0~,~\d\psi_{\a}=\Dd^2 L_{\a}+i\Dd^2\D_{\a}K~,~K=\bar{K}~.$
\item \underline{New-new-minimal}\cite{newnewm1}: $\d H_{\a\ad}=\D_{\a}\bar{L}_{\ad}-\Dd_{\ad}L_{\a}~,~$
$\d V=\D^{\a}\Dd^2L_{\a}-\Dd^{\ad}\D^2\bar{L}_{\ad}~,V$ is imaginary, linear. $V=\D^{\a}\phi_{\a}-\Dd^{\ad}\bar{\phi}_{\ad}~,~\Dd_\ad\phi_\a=0~,~\d\phi_{\a}=\Dd^2 L_{\a}+\Dd^2\D_{\a}K,\\ K=\bar{K}~.$ This formulation is known at the linearized level only.
\item \underline{Non-minimal}\cite{nonm1,nonm2}: $\d H_{\a\ad}=\D_{\a}\bar{L}_{\ad}-\Dd_{\ad}L_{\a}~,~\d\Gamma=\Dd^{\ad}\D^2\bar{L}_{\ad}+f(n)\Dd^2\D^{\a}L_{\a}~,$\\ $\Gamma$ is complex linear.
$\Gamma=\Dd^{\ad}\bar{\chi}_{\ad}~,~\d \chi_{\a}=\Dd^2 L_{\a}+\tfrac{1}{2}f^{*}(n)\D_{\a}\Dd^{\ad}\bar{L}_{\ad}+\D^{\b}L_{\b\a}$, with $L_{\b\a}=L_{\a\b}$ and $f(n)\neq\tfrac{1}{3},1,\infty$~.
\end{enumerate}
~~~~It is evident that most of the terms in (\ref{dSof}) can easily fit within the structure of the transformations of the supergravity superfields, therefore an interaction term can be found in order to get the invariant theory. To make this explicit, we go through the list of supergravities and identify the coupling terms. The general theme of this section is the following. By choosing a particular formulation of supergravity we choose a particular type of compensator. This translates to imposing constraints on $\N,\mathscr{M}$ and $\N_{\ad}$. Whether these constraints  can be satisfied (by fixing $W,\kappa,\lambda$) or not gives an indication to whether the theory can be coupled to this specific supergravity or not. As it was advertised, the superfield $X$ is the object that controls the outcome.

1.~\underline{\textbf{Old-minimal supergravity}}:\\
In order to be able to couple the theory to old-minimal supergravity, we must have
\bea{l}\n
\N=\D^2\Theta+\Dd^2\Xi~,~\text{for some}~\Theta,\Xi~,\sn\\
\mathscr{M}\neq \Dd^{\ad}Z_{\ad}~,~\text{for any}~Z_{\ad}~,\sn\\
c=0~,\sn\\
\kappa=0~,\sn\\
\Dd^{\bd}\bar{\L}_{\bd\ad}+\Dd_{\ad}\bar{\L}=0~.\sn
\eea
As a result, we get
{\allowdisplaybreaks
\bea{l}\n\label{omc}
X\neq -\tfrac{1}{2}(1+2\lambda)\S T_o -\tfrac{1}{2}\lambda^{*}\Sd\bar{T}_o
+\Dd^{\ad}Z_{\ad}+\tfrac{1}{2}\D^{\a}\bar{Z}_{\a}-\tfrac{1}{2}\D^2\bar{\Xi}-\tfrac{1}{2}\Dd^2\bar{\Theta}~,\sn\label{omXc}\\
W=-\tfrac{4}{3}X+\tfrac{5}{3}\bar{X}-\tfrac{2}{3}\S T_o+\tfrac{4}{3}\Sd\bar{T}_o+\D^2\left[\tfrac{4}{3}\Theta-\tfrac{2}{3}\bar{\Xi}\right]+\Dd^2\left[\tfrac{4}{3}\Xi-\tfrac{2}{3}\bar{\Theta}\right]~,\sn\label{omWc}\\
\Delta_{\ad}=-\D^2\bar{\L}_{\ad}~,\sn\label{omDadc}\\
\varphi=\lambda\Dd^2\D^{\a}\L_{\a}~,\sn\\
\Delta=-\Dd^{\ad}\D^2\bar{\L}_{\ad}+\lambda\Dd^2\D^{\a}\L_{\a}~.\sn\label{omDc}
\eea
}
This mean that for a given theory, we calculate $X$ according to (\ref{K}) and then we check if the constraint (\ref{omXc}) is satisfied. In other words, if there is a choice for the set $\lambda,\Theta,\Xi,Z_{\ad}$ such that equation (\ref{omXc}) is violated, then we can not couple the theory to old-minimal supergravity\footnote{There is the possibility that the coupling to supergravity happens only through superfield $H_{\a\ad}$ without any participation of the compensator and therefore we can not distinguish one supergravity from another. This corresponds to the case of conformal supergravity that will be examined separately.} for this choice. However, since the condition~(\ref{omXc}) is an exclusive one, it is fairly obvious that we can always couple the theory to old-minimal supergravity.

It would be useful to understand the meaning of (\ref{omc}). For equations (\ref{omXc}) and (\ref{omWc}) it is straightforward since they provide the condition for coupling to old-minimal formulation of supergravity and the appropriate improvement term we must use. Also, we have commented on the meaning of (\ref{omDc}) and how the existence of a non-zero $\Delta$ keeps $\Delta_{\ad}$ unconstrained in general. However, there is a little bit more in the meaning of (\ref{omDadc}). In constructing supergravities, one has to solve the anholonomy constraints for the supervielbeins in terms of prepotentials. Among the prepotentials, one introduces a real\footnote{after using the freedom of change of coordinates ($\mathcal{K}$-supergroup)} gauge supervector $H=(H_{\a\ad},\bar{H}_{\a},H_{\ad})$\footnote{look in \cite{ideas,GGRS} and references therein}with transformations
\bea{l}
\delta H_{\a\ad}\sim \Delta_{\a\ad}+\bar{\Delta}_{\a\ad}~,~
\delta H_{\ad}\sim\Delta_{\ad}-\bar{\Delta}_{\ad}\n~.
\eea
Since $\Delta_{\ad}$ is unconstrained we can use it to eliminate $H_{\ad}$. Equation (\ref{omDadc}) reflects the remaining symmetry in this $H_{\ad}$ fixed configuration, $\Delta_{\ad}=\bar{\Delta}_{\ad}=-\D^2\bar{\Lambda}_{\ad}$. 

Nevertheless, under conditions (\ref{omc}), equation (\ref{dSof}) takes the form
\bea{lll}
\d S_o &=\int d^8z ~&-\tfrac{1}{2}\left(\D^{\a}\bar{\L}^{\ad}-\Dd^{\ad}\L^\a\right)\N_{\a\ad}\n\\
& &+\Dd^2\D^{\a}\L_{\a}~\mathscr{M}+c.c.~.
\eea
In order to make contact with the linearized transformations of old-minimal supergravity, we must identify $\Lambda_{\a}$ with $L_{\a}$. Specifically, we must have $\Lambda_{\a}=\tfrac{1}{M}L_{\a}$, where $M$ is a mass scale. Of course, this is to be expected since the engineering dimensions of the two parameters do not match $[\Lambda_{\a}]=-\tfrac{3}{2}$ and $[L_{\a}]=-\tfrac{1}{2}$ but more importantly in order to be precise we need a parameter in the transformation (\ref{S-transf}) of $\S$ in order to keep track of the order up to which we work. The invariance of the action we attempt to construct will be valid up to linear order which translates up to $M^{-1}$ terms. In this case, the action is made invariant by adding the interaction terms
\bea{ll}
S_{int}=\int d^8z~&~\tfrac{1}{2M}H^{\a\ad}\N_{\a\ad}\n\label{int-om}\\
&-\tfrac{1}{M}\s\mathscr{M}+c.c.~.
\eea
From this expression, we can immediately read off that the supercurrents of the theory are $\mathscr{N_{\a\ad}}$ and $\mathscr{M}$.

2.~\underline{\textbf{New-minimal supergravity}}:\\
For coupling with new-minimal supergravity, we must have the following conditions:
\bea{l}\n\label{newmc}
\N=\mathscr{R}+[\D^2\Theta+
\Dd^2\Xi]~,~\text{for some}~\Theta,~\Xi,~\mathscr{R},~\text{with}~\mathscr{R}=\bar{\mathscr{R}}~,\sn\\
\mathscr{M}=\Dd^{\ad}Z_{\ad}~,~\text{for some}~Z_{\ad}~,\sn\\
c=0~,\sn\label{newmcc}\\
\kappa=0~,\sn\\
\Dd^{\bd}\bar{\L}_{\bd\ad}+\Dd_{\ad}\bar{\L}=0~\sn,
\eea
thus we get:
\bea{l}\n
X-\bar{X}=-(\tfrac{1+\lambda}{2})\S T_o + (\tfrac{1+\lambda^*}{2})\Sd\bar{T}_o+\tfrac{1}{2}\Dd^{\ad}Z_{\ad}-\tfrac{1}{2}\D^{\a}\bar{Z}_{\a}\sn\label{newmXc}\\
\hspace{10ex}+\tfrac{1}{2}\D^2\left[\Theta-\bar{\Xi}\right]-\tfrac{1}{2}\Dd^2\left[\bar{\Theta}-\Xi\right]~,\\
W=-\bar{X}-2\lambda^{*}\Sd\bar{T}_o+2\D^{\a}\bar{Z}_{\a}~,\sn\\
\Delta_{\ad}=-\D^2\bar{\L}_{\ad}~,\sn\\
\varphi=\lambda\Dd^2\D^{\a}\L_{\a}~,\sn\label{newmfc}\\
\Delta=-\Dd^{\ad}\D^2\bar{\L}_{\ad}+\lambda\Dd^2\D^{\a}\L_{\a}~.\sn\label{newmdc}
\eea
So only theories whose imaginary part of $X$ can be parametrized as in (\ref{newmXc}) for some $\lambda,\Theta,\Xi,Z_{\ad}$ can be coupled to new-minimal supergravity. In this case, equation~(\ref{dSof}) becomes
\bea{lll}
\d S_o &=\int d^8z ~&-\tfrac{1}{2}\left(\D^{\a}\bar{\L}^{\ad}-\Dd^{\ad}\L^\a\right)\N_{\a\ad}\n\label{dSnewm}\\
& &+\left(\D^{\a}\Dd^2\L_{\a}+\Dd^{\ad}\D^2\bar{\L}_{\ad}\right)\mathscr{R}~,
\eea
where
\bea{l}\n
\mathscr{R}=-(X+\bar{X})-(\tfrac{1+3\lambda}{2})\S T_o-(\tfrac{1+3\lambda^*}{2})\Sd \bar{T}_o+\tfrac{3}{2}\D^{\a}\bar{Z}_{\a}+\tfrac{3}{2}\Dd^{\ad}Z_{\ad}\\
\hspace{4.7ex}-\tfrac{1}{2}\D^2\left[\Theta+\bar{\Xi}\right]-\tfrac{1}{2}\Dd^2\left[\bar{\Theta}+\Xi\right]~,
\eea
and as a result the interaction terms we have to add are:
\bea{ll}
S_{int}=\int d^8z~&~\tfrac{1}{2M}H^{\a\ad}\N_{\a\ad}-\tfrac{1}{M}U\mathscr{R}\n\label{int-newm}~.
\eea
The supercurrents are $\N_{\a\ad},~\mathscr{R}$.

The above analysis can be generalized by relaxing the (\ref{newmcc}) condition. This can be done by exploiting the $K$ freedom in the transformation of the $\psi_{\a}$ superfield. We can modify the identification we do between $\L_{\a}$ and $L_{\a}$ in the following way:
\bea{l}\n
\Lambda_{\a}=\tfrac{1}{M}L_{\a}+\tfrac{i}{M}\D_{\a}K~.
\eea
Therefore, equation (\ref{dSnewm}) now takes the form
\bea{lll}\n
\d S_o &=\int d^8z ~&-\tfrac{1}{2M}\left(\D^{\a}\bar{L}^{\ad}-\Dd^{\ad}L^\a\right)\N_{\a\ad}\\
& &+\tfrac{1}{2M}K~\pa^{\a\ad}\N_{\a\ad}\\
& &+\tfrac{1}{M}\left(\D^{\a}\Dd^2 L_{\a}+\Dd^{\ad}\D^2\bar{L}_{\ad}\right)\mathscr{R}\\
& &+\tfrac{c}{M}\left(\D^2\bar{L}^{\ad}-i\D^2\Dd^{\ad}K\right)\N_{\ad}+c.c.~,
\eea
and the request for invariance leads to the introduction of the following interacting terms
\bea{lll}\n
S_{int}\n\label{int-newmg}
&=\int d^8z~&~\tfrac{1}{2M}H^{\a\ad}\N_{\a\ad}\\
& &+\tfrac{1}{M}\psi^{\a}\left(\D_{\a}\mathscr{R}-c^*\bar{\N}_{\a}\right)+c.c.~,
\eea
together with the constraint $\pa^{\a\ad}\N_{\a\ad}=0$.

3.~\underline{\textbf{New-new-minimal supergravity}}:\\
For new-new-minimal supergravity we should impose the conditions
\bea{l}\n
\N=i\mathscr{I}+[\D^2\Theta+
\Dd^2\Xi]~,~\text{for some}~\Theta,~\Xi,~\mathscr{I},~\text{with}~\mathscr{I}=\bar{\mathscr{I}}~,\sn\\
\mathscr{M}=\Dd^{\ad}Z_{\ad}~,~\text{for some}~Z_{\ad}~,\sn\\
c=0~,\sn\\
\kappa=0~,\sn\\
\Dd^{\bd}\bar{\L}_{\bd\ad}+\Dd_{\ad}\bar{\L}=0~,\sn
\eea
therefore we get:
\bea{l}\n
X+\bar{X}=-(\tfrac{1+3\lambda}{2})\S T_o -(\tfrac{1+3\lambda^*}{2})\Sd\bar{T}_o+\tfrac{3}{2}\Dd^{\ad}Z_{\ad}+\tfrac{3}{2}\D^{\a}\bar{Z}_{\a}\sn\label{newnewmXc}\\
\hspace{10ex}-\tfrac{1}{2}\D^2\left[\Theta+\bar{\Xi}\right]-\tfrac{1}{2}\Dd^2\left[\bar{\Theta}+\Xi\right]~,\\
W=-\bar{X}-2\lambda^{*}\Sd\bar{T}_o+2\D^{\a}\bar{Z}_{\a}~,\sn\\
\Delta_{\ad}=-\D^2\bar{\L}_{\ad}~,\sn\\
\varphi=\lambda\Dd^2\D^{\a}\L_{\a}~,\sn\\
\Delta=-\Dd^{\ad}\D^2\bar{\L}_{\ad}+\lambda\Dd^2\D^{\a}\L_{\a}~,\sn
\eea
which means that the theories that are allowed to couple to new-new-minimal supergravity are the ones for whom the real part of $X$ can be parametrized as in (\ref{newnewmXc}) for some $\lambda,\Theta,\Xi,Z_{\ad}$. The variation (\ref{dSof}) now takes the form
\bea{lll}
\d S_o &=\int d^8z ~&-\tfrac{1}{2}\left(\D^{\a}\bar{\L}^{\ad}-\Dd^{\ad}\L^\a\right)\N_{\a\ad}\n\label{dSnewnewm}\\
& &+i\left(\D^{\a}\Dd^2\L_{\a}+\Dd^{\ad}\D^2\bar{\L}_{\ad}\right)\mathscr{I}~,
\eea
where
\bea{l}\n
i\mathscr{I}=X-\bar{X}+(\tfrac{1+\lambda}{2})\S T_o-(\tfrac{1+\lambda^*}{2})\Sd \bar{T}_o+\tfrac{1}{2}\D^{\a}\bar{Z}_{\a}-\tfrac{1}{2}\Dd^{\ad}Z_{\ad}\\
\hspace{4.7ex}-\tfrac{1}{2}\D^2\left[\Theta-\bar{\Xi}\right]+\tfrac{1}{2}\Dd^2\left[\bar{\Theta}-\Xi\right]~,
\eea
and the interaction terms are:
\bea{ll}
S_{int}=\int d^8z~&~\tfrac{1}{2M}H^{\a\ad}\N_{\a\ad}-\tfrac{i}{M}V\mathscr{I}\n\label{int-newnewm}~.
\eea
The corresponding supercurrents are: $\N_{\a\ad},~i\mathscr{I}$.

However, we can do similar generalizations as we did for the previous case. In particular
\bea{l}\n
\Lambda_{\a}=\tfrac{1}{M}L_{\a}+\tfrac{1}{M}\D_{\a}K~,
\eea
and equation (\ref{dSnewnewm}) becomes
\bea{lll}\n
\d S_o &=\int d^8z ~&-\tfrac{1}{2M}\left(\D^{\a}\bar{L}^{\ad}-\Dd^{\ad}L^\a\right)\N_{\a\ad}\\
& &-\tfrac{1}{2M}K~\left[\D^{\a},\Dd^{\ad}\right]\N_{\a\ad}\\
& &+\tfrac{i}{M}\left(\D^{\a}\Dd^2 L_{\a}-\Dd^{\ad}\D^2\bar{L}_{\ad}\right)\mathscr{I}\\
& &+\tfrac{c}{M}\left(\D^2\bar{L}^{\ad}+\D^2\Dd^{\ad}K\right)\N_{\ad}+c.c.~.
\eea
The generalized interaction terms are
\bea{lll}\n\label{int-newnewmg}
S_{int}\n
&=\int d^8z~&~\tfrac{1}{2M}H^{\a\ad}\N_{\a\ad}\\
& &+\tfrac{1}{M}\phi^{\a}\left(i\D_{\a}\mathscr{I}-c^*\bar{\N}_{\a}\right)+c.c.~,
\eea
together with the constraint $\left[\D^{\a},\Dd^{\ad}\right]\N_{\a\ad}=0$.

4.~\underline{\textbf{Non-minimal supergravity}}:\\
Finally, we have the case of non-minimal supergravity. To couple the theory with non-minimal supergravity we must impose the following constraints:
\bea{l}\n
\N\neq \D^2\Theta+\Dd^2\Xi~,~\text{for any}~\Theta,~\Xi~,\sn\label{cf2}\\
\mathscr{M}=f\bar{\N}+\Dd^{\ad}Z_{\ad}~,~\text{for some}~f,~Z_{\ad}~,\sn\label{cf}\\
c\neq 0~,\sn\\
\kappa=0~,\sn\\
\Dd_{\ad}\bar{\L}=\tfrac{c}{2}f\Dd_{\ad}\D^{\a}\L_{\a}~,\sn
\eea
which gives:
\bea{l}\n
f\neq \tfrac{1}{3},1,\infty ~,\sn\\
X\neq \tfrac{1}{2}(-1-2\lambda+c)\S T_o -\tfrac{1}{2}\lambda^{*}\Sd\bar{T}_o
+\Dd^{\ad}Z_{\ad}+\tfrac{1}{2}\D^{\a}\bar{Z}_{\a}\sn\\
\hspace{5ex}+\tfrac{1}{2}\D^2\left[f\Theta+(2f-1)\bar{\Xi}\right]+\tfrac{1}{2}\Dd^2\left[(2f-1)\bar{\Theta}+f\Xi\right]~,\sn\label{nonmXc}\\
W=-\tfrac{4f^2}{(3f-1)(f-1)}X+\tfrac{(5f^2-1)}{(3f-1)(f-1)}\bar{X}-\tfrac{2f(\lambda+f-cf)}{(3f-1)(f-1)}\S T_o\sn\label{nonmW}\\
\hspace{5.5ex}+\tfrac{2(2f-1)(\lambda^{*}+f-c^*f)}{(3f-1)(f-1)}\Sd \bar{T}_o
+\tfrac{2f}{(3f-1)(f-1)}\Dd^{\ad}Z_{\ad}-\tfrac{2(2f-1)}{(3f-1)(f-1)}\D^{\a}\bar{Z}_{\a}~,\\
\Delta_{\ad}=(c-1)\D^2\bar{\L}_{\ad}+\tfrac{c}{2}f\Dd_{\ad}\D^{\a}\L_{\a}+\Dd^{\bd}\bar{\L}_{\bd\ad}\sn\label{nonmDadc}~,\\
\varphi=(\lambda-cf)\Dd^2\D^{\a}\L_{\a}\sn~,\\
\Delta=(c-1)\Dd^{\ad}\D^2\bar{\L}_{\ad}+\lambda\Dd^2\D^{\a}\L_{\a}\sn~.
\eea
Notice that $\Delta_{\ad}$ is not equal to $\bar\Delta_\ad$ anymore which means that we are not in the $H_{\ad}=0$ gauge. Instead $H_{\ad}$ transforms as the prepotential of the compensator $\Gamma$ and can be identified with it ($\delta H_{\ad}\sim\Delta_{\ad}-\bar{\Delta}_{\ad}=c\D^2\bar{\L}_{\ad}+\tfrac{c}{2}f\Dd_{\ad}\D^{\a}\L_{\a}+\Dd^{\bd}\bar{\L}_{\bd\ad}$, precisely the transformation of the prepotential of the complex linear compensator as discussed before).

Furthermore, the constraints on the parameter $f$ are precisely the constraints imposed in non-minimal supergravity. The $f\neq \tfrac{1}{3}$ and $f\neq 1$ constraints emerge from the self-consistency of equation (\ref{cf}) whereas the exclusion of  $f=\infty$ emerge from the consistency of (\ref{cf}) with (\ref{cf2}). The meaning of these constraints can be understood through equation (\ref{IT}). It is easy to show that for $f=1$ we fall back to the new-minimal configuration and for $f=\tfrac{1}{3}$ we go to the new-new-minimal case. This is also indicated from the $f=1$ and $f=\tfrac{1}{3}$ limits of $(3f-1)(f-1)W$ which due to (\ref{nonmW}) makes contact with equations (\ref{newmXc}~,~\ref{newnewmXc}). The $f=\infty$ case can also be shown to correspond to old-minimal supergravity. Historically, $f$ was parametrized by a number $n$ such that $f(n)=\tfrac{n+1}{3n+1}$ with $n\neq \infty,0,-\tfrac{1}{3}$ which matches the exclusion of $f=\tfrac{1}{3},1,\infty$.

Moreover, there is a choice of parameters that makes $\Delta$ vanish ($c=1, \lambda=0$). Whenever such a choice is compatible with (\ref{nonmXc}) it leads to a group action without scaling type terms.

The variation of the action for this case takes the form:
\bea{lll}
\d S_o &=\int d^8z ~&-\tfrac{1}{2}\left(\D^{\a}\bar{\L}^{\ad}-\Dd^{\ad}\L^\a\right)\N_{\a\ad}\\
& &+\left(\D^{\a}\Dd^2\L_{\a}+f\D^2\Dd^{\ad}\bar{\L}_{\ad}\right)\N+c.c.\n\\
& &+c\left(\D^2\bar{\L}^{\ad}
-\tfrac{1}{c}\Dd_{\bd}\bar{\L}^{\bd\ad}+\tfrac{f}{2}\Dd^{\ad}\D^{\a}\L_{\a}\right)\N_{\ad}+c.c.~,
\eea
therefore, the interactions we have to introduce are ($\L_{\a}=\tfrac{1}{M}L_{\a}~,~\L_{\b\a}=-\tfrac{c}{M}L_{\b\a}$)
\bea{ll}
S_{int}=\int d^8z~&~\tfrac{1}{2M}H^{\a\ad}\N_{\a\ad}\n\label{int-nonm}\\
&+\tfrac{1}{M}\chi^{\a}\mS_{\a}+c.c.~,
\eea
where $\mS_{\a}=\D_{\a}\N-c^*\bar{\N}_{\a}$.
From this we conclude that the corresponding supercurrents are $\N_{\a\ad}$ and $\mS_{\a}$. An interesting observation is that if $c\neq 0$ then, the $\S$ theory has an interaction term directly with the prepotential of $\Gamma$.

5.~\underline{\textbf{Conformal supergravity}}:\\
For completeness, we examine the special case where the complex linear theory interacts to supergravity only through the $H_{\a\ad}$ superfield. This corresponds to the coupling of the theory to conformal supergravity. We must have: 
\bea{l}\n
\N=\D^2\Theta+\Dd^2\Xi~,~\text{for some}~\Theta,\Xi~,\sn\\
\mathscr{M}= \Dd^{\ad}Z_{\ad}~,~\text{for some}~Z_{\ad}~,\sn\\
c=0~,\sn\\
\kappa=0~,\sn\\
\Dd^{\bd}\bar{\L}_{\bd\ad}+\Dd_{\ad}\bar{\L}=0\sn~.
\eea
Therefore this can happen for the theories where $X$ takes the special form
\bea{l}\n\label{dgXc}
X=-\tfrac{1+2\lambda}{2}\S T_o-\tfrac{\lambda^{*}}{2}\Sd\bar{T}_o+\Dd^{\ad}Z_{\ad}+\tfrac{1}{2}\D^{\a}\bar{Z}_{\a}-\tfrac{1}{2}\D^2\bar{\Xi}-\tfrac{1}{2}\Dd^2\bar{\Theta}~,
\eea
and the coupling to superconformal supergravity is:
\bea{ll}\n\label{int-dg}
S_{int}=\int d^8z~&~\tfrac{1}{2M}H^{\a\ad}\N_{\a\ad}~.
\eea
%
%
%
%
\section{Bianchi Identities and Conservation equations}\label{consEq}
~~~~The invariance of the full action will be expressed by a set of Bianchi identities. By taking the on-shell limit of these identities we recover the conservation equations of the supercurrents.
Let's assume that we have a theory of complex linear superfields coupled to one of the supergravities, $S=S[\S,\Sd,H_{\a\ad},C]$ where $C$ is the compensator of the specific supergravity. Therefore, the variation of the action under the linearized transformations is:
\bea{ll}
\d S=\int d^8z~&~ \left\{\Delta^{\a}\D_{\a}\S+\Delta^{\ad}\Dd_{\ad}\S+i\Delta^{\a\ad}\pa_{\a\ad}\S+\Delta\S\right\}
\mathcal{T}+c.c.\\
&+\left\{\D^{\a}\bar{L}^{\ad}-\Dd^{\ad}L^{\a}\right\}\mathcal{T}_{\a\ad}\n\\
&+\d C~\mathcal{T}_{C}+c.c.~,
\eea
with $\mathcal{T}=\tfrac{\d S}{\d \S},~\mathcal{T}_{\a\ad}=\tfrac{\d S}{\d H^{\a\ad}},~\mathcal{T}_{C}=\tfrac{\d S}{\d C}$ and
\bea{l}\n
\Delta_{\a}=-\tfrac{1}{M}\Dd^2 L_{\a}~,~\Delta_{\ad}=\tfrac{p_1}{M}\D^2\bar{L}_{\ad}+\tfrac{p_2}{M}\Dd^{\bd}\bar{L}_{\bd\ad}+\tfrac{p_3}{M}\Dd_{\ad}\D^{\a}L_{\a}~,\\
\Delta_{\a\ad}=\tfrac{1}{M}\Dd_{\ad}L_{\a}~,~\Delta=\tfrac{p_1}{M}\Dd^{\ad}\D^2\bar{L}_{\ad}+\tfrac{\lambda}{M}\Dd^2\D^{\a}L_{\a}~,
\eea
where $p_1=-1,~p_2=p_3=0$ for the minimal cases and $p_1=c-1,~p_2=-c,~p_3=\tfrac{c}{2}f$ for the non-minimal case. Hence
\bea{ll}
\d S=\int d^8z~&~L^{\a}\left\{-\tfrac{1}{M}\Dd^2\left[\D_{\a}\S~\mathcal{T}+
p^{*}_{1}\Sd\D_{\a}\bar{\mathcal{T}}\right]+\tfrac{p_3}{M}\D_{\a}\Dd^{\ad}\left[\Dd_{\ad}\S~\mathcal{T}\right]\right.\\
&~~~~~~~\left.+\tfrac{i}{M}\Dd^{\ad}\left[\pa_{\a\ad}\S~\mathcal{T}\right]-\tfrac{\lambda}{M}\D_{\a}\Dd^2\left[\S~\mathcal{T}\right]-\Dd^{\ad}\mathcal{T}_{\a\ad}\right\}+c.c.\n\\
&+p^{*}_{2}L^{\b\a}\left\{-\tfrac{1}{2M}\D_{(\b}\left[\D_{\a)}\Sd~\bar{\mathcal{T}}\right]\right\}+c.c.\\
&+\d C~\mathcal{T}_{C}+c.c. ~.
\eea
Now we will use the above to derive the Bianchi identities for all the previously discussed supergravities and the theory $S[\S,\Sd,H_{\a\ad},C]=S_o[\S,\Sd]+S_{int}[\S,\Sd,H_{\a\ad},C]$.

1.~\underline{\textbf{Old-minimal supergravity}}:\\
In this case, $C$ is a chiral superfield $\s$ with linearized transformation
$\delta\s=\Dd^2\D^{\a}L_{\a}$. So the variation of the action is
\bea{ll}
\d S=\int d^8z~&~L^{\a}\left\{-\tfrac{1}{M}\Dd^2\left[\D_{\a}\S~\mathcal{T}-\Sd\D_{\a}\bar{\mathcal{T}}\right]+\tfrac{i}{M}\Dd^{\ad}\left[\pa_{\a\ad}\S~\mathcal{T}\right]-\tfrac{\lambda}{M}\D_{\a}\Dd^2\left[\S~\mathcal{T}\right]\right.\\
&~~~~~~~\left.\vphantom{\tfrac{1}{M}}-\Dd^{\ad}\mathcal{T}_{\a\ad}-\D_{\a}\Dd^2\mathcal{T}_{\s}\right\}+c.c.~.\n
\eea
Therefore the invariance of the action provides the following Bianchi identity
\bea{l}\n
\hspace{-3ex}\Dd^2\left\{\D_{\a}\S~\mathcal{T}-\Sd\D_{\a}\bar{\mathcal{T}}\right\}-
\Dd^{\ad}\left\{i\pa_{\a\ad}\S~\mathcal{T}-M\mathcal{T}_{\a\ad}\right\}+\D_{\a}\Dd^{2}\left\{\lambda\S\mathcal{T}+M\mathcal{T}_{\s}\right\}=0~.
\eea
However, due to (\ref{int-om}) we have that
\bea{l}\n
\mathcal{T}_{\a\ad}=\tfrac{1}{2M}\N_{\a\ad}~,~\mathcal{T}_{\s}=-\tfrac{1}{M}\mathscr{M}~,
\eea
and by going on-shell, using the $\S$ equation of motion $\left(\Dd_{\ad}\mathcal{T}=0\right)$ we get the following conservation equation:
\bea{l}\n\label{ce-om}
\Dd^{\ad}\N_{\a\ad}=2\D_{\a}\Dd^{2}\mathscr{M}~.
\eea
Of course, this is the well-known Ferrara-Zumino multiplet \cite{sc1}.

2.~\underline{\textbf{New-minimal supergravity}}:\\
When coupling to new-minimal, we get the Bianchi identity
\bea{l}\n
\Dd^2\left\{\D_{\a}\S~\mathcal{T}-\Sd\D_{\a}\bar{\mathcal{T}}\right\}-
\Dd^{\ad}\left\{i\pa_{\a\ad}\S~\mathcal{T}-M\mathcal{T}_{\a\ad}\right\}\\
+\lambda\D_{\a}\Dd^{2}\left\{\S\mathcal{T}\right\}+M\Dd^2\D_{\a}\mathcal{T}_{U}=0~.
\eea
Going on-shell ($\Dd_{\ad}\mathcal{T}=0$) and using that $\mathcal{T}_{\a\ad}=\tfrac{1}{2M}\N_{\a\ad}~,~\mathcal{T}_{U}=-\tfrac{1}{M}\mathscr{R}$ we get the conservation equation
\bea{l}\n\label{newmCE}
\Dd^{\ad}\N_{\a\ad}=2\Dd^{2}\D_{\a}\mathscr{R}~.
\eea
This is the $\mathscr{R}$-multiplet \cite{sc2,GGRS}. The structure of this conservation equation together with the reality of $\mathscr{R}$ results in the spacetime conservation of the entire supercurrent $\N_{\a\ad}$: $\pa^{\a\ad}\N_{\a\ad}=0$. This corresponds to the fact that there is an extra U(1) symmetry due to R-symmetry.
R-symmetry rotates the superspace fermionic coordinates $\theta\to e^{ia}\theta,~\bar{\theta}\to e^{-ia}\bar{\theta}$ and if superfield $\S$ has a well defined R-charge $q$ it transforms $\S\to e^{iqa}\S$. It is straightforward to check that at the linear limit, this transformation of $\S$ fits exactly in the (\ref{dS}) form with
\bea{ll}\n
\Delta_{\a}=-ia\theta_{\a}~,~&\Delta_{\a\ad}=ia\theta_{\a}\bar{\theta}_{\ad}~,\\
\Delta_{\ad}=ia\bar{\theta}_{\ad}~,~&\Delta=iaq~,
\eea
which can be checked to satisfy all of the (\ref{const}) constraints.
Also, in the language of (\ref{solconst}) R-symmetry transformation corresponds to the choice
$\Lambda_{\a}=-ia~\D_{\a}\left[\theta^2\bar{\theta}^2\right]$ with $\varphi=i(q+2)a$ and the parametrization $q=-2(1+\lambda)$.
By plugging this value of $\L_{\a}$ into (\ref{dSof}) and demand invariance under R-symmetry, we get the following requirements:
\bea{l}\n
\N-\bar{\N}=\text{chiral}+\text{antichiral}~,\sn\label{r1}\\
\mathscr{M}=\text{complex linear}~,\sn\label{r2}\\
c=\kappa=0~,\sn\label{r3}\\
\Dd^{\bd}\bar{\L}_{\bd\ad}+\Dd_{\ad}\bar{\L}=0~,\sn\label{r4}\\
\pa^{\a\ad}\N_{\a\ad}=0~.\sn\label{r5}
\eea
We can check that (\ref{r1},~\ref{r2},~\ref{r3},~\ref{r4}) are exactly the requirements for coupling the theory to new-minimal supergravity (\ref{newmc}) and the conservation of $\N_{\a\ad}$ (\ref{r5}) is consistent with the superspace conservation equation for the new-minimal case (\ref{newmCE}).

Now, if we consider the slightly more general treatment of the new-minimal case~(\ref{int-newmg}) then we get the following:
\bea{l}\n
\Dd^{\ad}\N_{\a\ad}=2\Dd^2\D_{\a}\mathscr{R}-2c^*\Dd^2\bar{\N}_{\a}\sn~,\\
c^*\D^{\g}\Dd^2\bar{\N}_{\g}-c\Dd^{\gd}\D^2\N_{\gd}=0\sn~.
\eea
For the expansion order we are working (up to $1/M$), $\N_{\ad}$ (\ref{Nad}) can be written on-shell as $\N_{\ad}=\Dd_{\ad}\left[\S T_{o}\right]$, hence we get that
\bea{l}\n
\Dd^{\ad}\N_{\a\ad}=2\Dd^2\D_{\a}\left[\mathscr{R}-\tfrac{c^*}{2}\Sd \bar{T}_{o}-\tfrac{c}{2}\S T_{o}\right]\sn~,\label{d8}\\
c^*\Sd \bar{T}_{o}-c\S T_{o}=\D^2 P - \Dd^2 \bar P \sn~,\label{d9}
\eea 
where $P$ is an arbitrary superfields. A non-trivial solution ($c\neq 0$) of condition~(\ref{d9}) imposes severe constraints on the starting action. A class of such solution are the K\"{a}hler sigma models which are polynomials of $\S\Sd$. However, from the point of view of higher derivative theories the non-trivial solutions will correspond to the $c=0$. This will be the class of solutions that we will consider here.

3.~\underline{\textbf{New-new-minimal supergravity}}:\\
Similarly to the previous case, the conservation equation of the new-new-minimal supercurrent is \cite{sNp2,sc3}
\bea{l}\n
\Dd^{\ad}\N_{\a\ad}=2i\Dd^{2}\D_{\a}\mathscr{I}~.
\eea
As in the new-minimal case, the Bianchi identities that originate from the more general (\ref{int-newnewmg}) treatment give slightly more abstract conditions. However, for the same reasons as in the new-minimal discussion we will consider only the $c=0$ class of solutions which corresponds to the analysis presented in the previous section. 
 
4.~\underline{\textbf{Non-minimal supergravity}}:\\
For non-minimal supergravity, $C$ is an unconstrained, spinorial superfield $\chi_{\a}$ with\\
$\d\chi_{\a}=\Dd^2 L_{\a}+\tfrac{1}{2}f\D_{\a}\Dd^{\ad}\bar{L}_{\ad}+\D^{\b}L_{\b\a}$.
The corresponding Bianchi identities are:
\bea{l}\n
\Dd^2\left\{\D_{\a}\S~\mathcal{T}+(c^*-1)\Sd\D_{\a}\bar{\mathcal{T}}-M\mathcal{T}_{\a}\right\}-
\Dd^{\ad}\left\{i\pa_{\a\ad}\S~\mathcal{T}-M\mathcal{T}_{\a\ad}\right\}\sn\\
-\tfrac{f}{2}\D_{\a}\Dd^{\ad}\left\{c\Dd_{\ad}\S~\mathcal{T}+M\bar{\mathcal{T}}_{\ad}\right\}
+\lambda\D_{\a}\Dd^{2}\left\{\S\mathcal{T}\right\}=0~,\\
c^*\D_{(\b}\left\{\D_{\a)}\Sd~\bar{\mathcal{T}}\right\}+M\D_{(\b}\mathcal{T}_{\a)}=0\sn~.
\eea
Therefore the conservation equations we get are
\bea{l}\n
\Dd^{\ad}\N_{\a\ad}=2\Dd^2\mS_{\a}+f\D_{\a}\Dd^{\ad}\bar{\mS}_{\ad}\sn~,\\
\D_{(\b}\mS_{\a)}=c^*\D_{(\b}\bar{\N}_{\a)}=0\sn~.
\eea
These can be re-written in the form
\bea{l}\n\label{ce-nonm}
\Dd^{\ad}\N_{\a\ad}=2\Dd^2\D_{\a}\mS+2f\D_{\a}\Dd^2\bar{\mS}~,
\eea
where $\mS_{\a}=\D_{\a}\mS$ and $\mS=\N-c^*\Sd\bar{T}_o$. However $\mS$ is not uniquely 
defined, it has a gauge freedom $\delta\mS=\D^2\mathscr{F}$. This is the non-minimal multiplet \cite{sc3}.

5.~\underline{\textbf{Conformal supergravity}}:\\
In this case, there is no compensator and the conservation equation
for the supercurrent has a very simple form
\bea{l}
\Dd^{\ad}\N_{\a\ad}=0\n~.
\eea

\bigskip
To make contact with the notation in \cite{sc2}, all the above conservation equations can be organized in the system
\bea{l}\n
\Dd^{\ad}\mathcal{S}_{\a\ad}=\D_{\a}\mathcal{Z}+\mathcal{X}_{\a}+\mathcal{X}^\prime_{\a}~,\\
\Dd_{\ad}\mathcal{Z}=0~,\\
\Dd_{\ad}\mathcal{X}_{\a}=0~,~\D^{\a}\mathcal{X}_{\a}-\Dd^{\ad}\bar{\mathcal{X}}_{\ad}=0~,\\
\Dd_{\ad}\mathcal{X}^\prime_{\a}=0~,~\D^{\a}\mathcal{X}^\prime_{\a}+\Dd^{\ad}\bar{\mathcal{X}}^\prime_{\ad}=0~,
\eea
with the following correspondence.
\begin{enumerate}[I]
\item \underline{{Old-minimal}}: $\mathcal{S}_{\a\ad}=\N_{\a\ad}~,~\mathcal{Z}=2\Dd^2\mathscr{M}~,~\mathcal{X}_{\a}=0~,~\mathcal{X}^\prime_{\a}=0$.
\item \underline{{New-minimal}}: $\mathcal{S}_{\a\ad}=\N_{\a\ad}~,~\mathcal{Z}=0~,~\mathcal{X}_{\a}=2\Dd^2\D_{\a}\mathscr{R}~,~\mathcal{X}^\prime_{\a}=0$.
\item \underline{{New-new-minimal}}: $\mathcal{S}_{\a\ad}=\N_{\a\ad}~,~\mathcal{Z}=0~,~\mathcal{X}_{\a}=0~,~\mathcal{X}^\prime_{\a}=2i\Dd^2\D_{\a}\mathscr{I}$.
\item \underline{{Non-minimal}}: $\mathcal{S}_{\a\ad}=\N_{\a\ad}~,~\mathcal{Z}=2f\Dd^2\bar{\mS}~,~\mathcal{X}_{\a}=\Dd^2\D_{\a}(\mS+\bar{\mS})~,~\mathcal{X}^\prime_{\a}=\Dd^2\D_{\a}(\mS-\bar{\mS})$.
\item \underline{{Conformal}}: $\mathcal{S}_{\a\ad}=\N_{\a\ad}~,~\mathcal{Z}=0~,~\mathcal{X}_{\a}=0~,~\mathcal{X}^\prime_{\a}=0$.
\end{enumerate}
For the case of non-minimal supergravity, we have seen that the parameter $f$ is not arbitrary because for specific values of it the interaction terms can be recast in terms of the minimal descriptions. This can be seen independently from the conservation equations (\ref{ce-nonm}) which can be rewritten in the following manner:
\bea{l}\n
\Dd^{\ad}\left\{\N_{\a\ad}+2f\D_{\a}\Dd_{\ad}\bar{\mS}-2f\Dd_{\ad}\D_{\a}\mS\right\}=2\Dd^2\D_{\a}\left\{(1-2f)\mS-f\bar{\mS}\right\}~.
\eea
Hence for $f=1$ we get
\bea{l}\n
\Dd^{\ad}\left\{\N_{\a\ad}+2\D_{\a}\Dd_{\ad}\bar{\mS}-2\Dd_{\ad}\D_{\a}\mS\right\}=-2\Dd^2\D_{\a}\left\{\mS+\bar{\mS}\right\}~,
\eea
corresponding to the conservation equation of new-minimal supergravity with\\
$\mathcal{S}_{\a\ad}=\N_{\a\ad}+2\D_{\a}\Dd_{\ad}\bar{\mS}-2\Dd_{\ad}\D_{\a}\mS~,~\mathcal{X}_{\a}=-2\Dd^2\D_{\a}(\mS+\bar{\mS})$.\\
For $f=\frac{1}{3}$ we get 
\bea{l}\n
\Dd^{\ad}\left\{\N_{\a\ad}+\tfrac{2}{3}\D_{\a}\Dd_{\ad}\bar{\mS}-\tfrac{2}{3}\Dd_{\ad}\D_{\a}\mS\right\}=\tfrac{2}{3}\Dd^2\D_{\a}\left\{\mS-\bar{\mS}\right\}~,
\eea
corresponding to the conservation equation of new-new-minimal supergravity with
$\mathcal{S}_{\a\ad}=\N_{\a\ad}+\tfrac{2}{3}\D_{\a}\Dd_{\ad}\bar{\mS}-\tfrac{2}{3}\Dd_{\ad}\D_{\a}\mS~,~\mathcal{X}^\prime_{\a}=\tfrac{2}{3}\Dd^2\D_{\a}(\mS-\bar{\mS})$. This is an elegant alternative to the usual argument involving improvement terms. From this point of view, there is no need for improvement terms and the algebra provides the exact redefinitions that need to be done in order to match the two formulations.
For the $f=\infty$ limit, we do not have to do anything since from (\ref{ce-nonm}) it is obvious that the term corresponding to old-minimal coupling dominates.

The superspace conservation equations include all information about the superdiffeomorphism invariance of the theory. To investigate the properties of the theory under different parts of the superdiffeomorphism group we project the conservation equations into components and discover the corresponding conserved currents. This procedure is straight forward and we will demonstrate it for equation (\ref{ce-nonm}) since all other irreducible configurations can be extracted from this one. We define the various components of a superfield through the action of covariant derivatives on the superfield and setting  the $\theta$ coordinates to zero. In this way by acting with derivatives from the left of (\ref{ce-nonm}) and projecting we extract the various component equations. The results for the independent equations are\footnote{The various components are labeled by the name of the superfield they come from and their position $(n,m)$ in its $\theta$ expansion. For example, $\Phi^{(0,0)}$ is the $\theta$ independent term of superfield $\Phi$, $\Phi^{(0,1)}_{\ad}$ is the $\bar{\theta}$ component and $\Phi^{(1,1)}_{\a\ad}$ is its $\theta\bar{\theta}$ component. Components with more than one index of the same type can be decomposed into symmetric (S) and anti-symmetric (A) pieces as $\Phi^{(S)}_{\b\a}=\Phi_{(\b\a)}$~,~$\Phi^{(A)}=C^{\b\a}\Phi_{\b\a}$.}:
\bea{ll}\n
(0,0):~&\N^{(A)(0,1)}_{\a}=
2\[\mS^{(1,2)}_{\a}+f\bar{\mS}^{(1,2)}_{\a}\]
+ i\pa_{\a}{}^{\gd}\[\mS^{(0,1)}_{\gd}-f\bar{\mS}^{(0,1)}_{\gd}\]~,\sn\\
(1,0):~&\N^{(S,A)(1,1)}_{\b\a}=
-\tfrac{i}{2}\pa_{(\b}{}^{\gd}\N^{(0,0)}_{\a)\gd}
+ 2 i \pa_{(\b}{}^{\gd}\mS^{(1,1)}_{\a)\gd} 
~,\sn\\
&\N^{(A,A)(1,1)}=
\tfrac{i}{2}\pa^{\g\gd}\N^{(0,0)}_{\g\gd}-2if\pa^{\g \gd}\bar{\mS}^{(1,1)}_{\g \gd}\sn\\
&\hspace{12ex}-4\[\mS^{(2,2)}+f\bar{\mS}^{(2,2)}\] 
+\Box\[\mS^{(0,0)}-f\bar{\mS}^{(0,0)}\]~,\\
(0,1):~&\N^{(0,2)}_{\a\ad}=-2if\pa_{\a\ad}\bar{\mS}^{(0,2)} ~,\sn\\
(1,1):~&\N^{(S)(1,2)}_{\b\a\ad}=
\tfrac{i}{4} \pa_{(\b}{}^{\gd}\N^{(S)(0,1)}_{\a)\ad\gd}
+\tfrac{i}{2}\pa_{(\b\ad}\[\mS^{(1,2)}_{\a)}-3f\bar{\mS}^{(1,2)}_{\a)}\]\sn\\
&\hspace{10ex}-\tfrac{1}{4}\pa_{(\b\ad}\pa_{\a)}{}^{\gd}\[\mS^{(0,1)}_{\gd}+3f\bar{\mS}^{(0,1)}_{\gd}\]~,\\
&\N^{(A)(1,2)}_{\ad}=
-\tfrac{i}{4}\pa^{\g\gd}\N^{(S)(0,1)}_{\g\ad\gd}
-\tfrac{i}{2}\pa^{\g}{}_{\ad}\[\mS^{(1,2)}_{\g}+5f\bar{\mS}^{(1,2)}_{\g}\]\sn\label{N1}\\
&\hspace{11ex}-\tfrac{1}{4}\Box\[\mS^{(0,1)}_{\ad}-5f\bar{\mS}^{(0,1)}_{\ad}\]~,\\
(2,0):~&\N^{(A)(1,2)}_{\ad}=
\tfrac{i}{4}\pa^{\g\gd}\N^{(S)(0,1)}_{\g\ad\gd}
+\tfrac{i}{2}\pa^{\g}{}_{\ad}\[\mS^{(1,2)}_{\g}+(f-4)\bar{\mS}^{(1,2)}_{\g}\]\sn\label{N2}\\
&\hspace{10ex}+\tfrac{1}{4}\Box\[\mS^{(0,1)}_{\ad}-(f+4)\bar{\mS}^{(0,1)}_{\ad}\]~,\\
(2,1):~&\N^{(2,2)}_{\a\ad}= 
-\tfrac{i}{8}\pa^{\g\gd}\N^{(S,S)(1,1)}_{\g\a\gd\ad}+\tfrac{1}{16}\pa_{\a\ad}\pa^{\g\gd}\N^{(0,0)}_{\g\gd}-\tfrac{1}{4}\Box\N^{(0,0)}_{\a\ad}\sn\\
&\hspace{9ex}+\tfrac{i}{2}\pa_{\a\ad}\[S^{(2,2)}-3f\bar{\mS}^{(2,2)}\]
+\tfrac{1}{4}\pa_{\a\ad}\pa^{\g\gd}\[\mS^{(1,1)}+(3f-1)\bar{\mS}^{(1,1)}_{\g\gd}\]\\
&\hspace{9ex}-\tfrac{1}{2}\Box\[\mS^{(1,1)}_{\a\ad}-\bar{\mS}^{(1,1)}_{\a\ad}\]
-\tfrac{i}{8}\pa_{\a\ad}\Box\[\mS^{(0,0)}+3f\bar{\mS}^{(0,0)}\]~.
\eea
At this point there are a few interesting observations we can make. First of all due to the reality of
$\N^{(2,2)}_{\a\ad}$ we get the conservation equation
\bea{l}\n
\pa^{\g\gd}\N^{(S,S)(1,1)}_{\g\a\gd\ad}+i(2-3f)\pa_{\a\ad}\pa^{\g\gd}\[\mS^{(1,1)}_{\g\gd}-\bar{\mS}^{(1,1)}_{\g\gd}\]-4i\Box\[\mS^{(1,1)}_{\a\ad}-\bar{\mS}^{(1,1)}_{\a\ad}\]\\
\hspace{4ex}
-2(1-3f)\pa_{\a\ad}\[\mS^{(2,2)}+\bar{\mS}^{(2,2)}\]
+\tfrac{1}{2}(1+3f)\pa_{\a\ad}\Box\[\mS^{(0,0)}+\bar{\mS}^{(0,0)}\]=0~,
\eea
which is the conservation equation for the energy-momentum tensor associated with spacetime translations. Secondly, the consistency of equations (\ref{N1},~\ref{N2}) give the conservation equation of the fermionic current.
\bea{l}\n\label{cfc}
\hspace{-3ex}\pa^{\g\gd}\N^{(S)(0,1)}_{\g\ad\gd}+2\pa^{\g}{}_{\ad}\[\mS^{(1,2)}_{\g}+(3f-2)\bar{\mS}^{(1,2)}_{\g}\]-i\Box\[\mS^{(0,1)}_{\ad}-(3f+2)\bar{\mS}^{(0,1)}_{\ad}\]=0\,,
\eea
which is the conservation equation for the supercurrent associated with supersymmetry transformations. The third observation is the reality of $\N^{(A,A)(1,1)}$ which leads to
\bea{l}\n
i\pa^{\g\gd}\N^{(0,0)}_{\g\gd}+(f+1)\Box\[\mS^{(0,0)}-\bar{\mS}^{(0,0)}\]-2if\pa^{\g\gd}\[\mS^{(1,1)}_{\g\gd}+\bar{\mS}^{(1,1)}_{\g\gd}\]\\
+4(f-1)\[\mS^{(2,2)}-\bar{\mS}^{(2,2)}\]=0~.
\eea
The presence of the last, algebraic term causes the failure of the conservation of a vector current. However, for the special case of $f=1$
\footnote{As we have seen $f=1$ corresponds to new-minimal supergravity and therefore the theory has R-symmetry}
this obstacle is removed and the conserved current corresponds to the  $U(1)_R$ current of new-minimal supergravity.
%
%
%
%
\section{Examples}
~~~~In this section, we apply all the above-derived results to three specific examples of increasing complexity. The first example is the simplest possible one, the almost free theory

A.~\underline{\textbf{Almost free theory}}:\\
We consider the Lagrangian
\bea{l}
\mathcal{L}_{o}=-\S\Sd+g\S^2+g^*\Sd^2~.
\eea
For this system we can easily find the $X$ and $\Omega_{\b\a\ad}$ superfields
\bea{l}
X =-\tfrac{1}{2}\S\Sd +\tfrac{g}{2}\S^2 +\tfrac{g^*}{2}\Sd^2~,~\Omega_{\b\a\ad}=0~, 
\eea
and we can determine to which supergravity this system can be coupled to.
\begin{enumerate}
\item \underline{Old-minimal}: 
It is easy to check that equation (\ref{omXc}) is always satisfied, which means that this theory can always be coupled to old-minimal supergravity with supercurrents
where
\bea{l}\n
\N_{\a \ad}=~ i \pa_{\a \ad} \Sd \S   + \tfrac{1}{3} \D_{\a} \Dd_{\ad} \left ( \Sd \S  -4 g \S^2 + 2 g^* \Sd^2 \right )  + c.c.~, \sn\\ 
\mathscr{M}=~ - \left ( \tfrac{2}{3} + \lambda \right ) \S \Sd + \left ( \tfrac{5}{3} + 2 \lambda \right ) g \S^2  - \tfrac{1}{3} g^* \Sd^2~.\sn
\eea
It is straightforward to check conservation equation (\ref{ce-om}) is satisfied for the above supercurrents. 
\item \underline{New-minimal}:
In order to couple this theory with new-minimal supergravity, due to equation (\ref{newmXc}) we must have $g\neq 0,~\lambda=-1$ and the supercurrents are
\bea{ll}\n
\N_{\a \ad} =~ i \pa_{\a \ad} \Sd \S  +  \D_{\a} \Dd_{\ad} \left ( \Sd \S  - 2 g \S^2 \right )  + c.c.~,\sn\\ 
\mathscr{R} =~ -\S \Sd + g \S^2 + g^* \Sd^2~. \sn
\eea
Notice that the fixed value of $\lambda = -1$ corresponds to an explicit zero R-charge of $\S$ which is the only value for which the action is R-symmetry invariant.
\item \underline{New-new-minimal}: To couple with new-new-minimal supergravity, we must have (\ref{newnewmXc}) $g\neq 0~, \lambda=-2/3$ and
\bea{ll}\n
\N_{\a \ad}=~ i \pa_{\a \ad} \Sd \S  +  \D_{\a} \Dd_{\ad} \left ( \tfrac{1}{3}  \Sd \S  - \tfrac{2}{3} g \S^2 \right )  + c.c. ~,\sn\\ 
i\mathscr{I}=~\tfrac{1}{3} \left( g \S^2 - g^* \Sd^2  \right )~.\sn 
\eea
\item \underline{Non-minimal}: Due to (\ref{nonmXc}) we find that this theory can always be coupled to non-minimal supergravity. The supercurrents for this case are:
\bea{l}\n
\N_{\a \ad}= i \pa_{\a \ad} \Sd \S+\Dd_{\ad} \D_{\a} \left \{ \left ( \gamma_1 + \tfrac{1}{2} \right )  \Sd \S  +  \left ( {\gamma_2} + \tfrac{1}{2} \right ) g \S^2 +   \left ( {\gamma_3} - \tfrac{3}{2} \right ) g^* \Sd^2 \right \}\nonumber\\
\hspace{6ex}+ c.c.~,\sn\\
\mS_{\a}=~\D_{\a} \bigg [ \left ( \tfrac{5}{4}  +  {\gamma_1} + \tfrac{1}{2} \bar {\gamma}_1  - c^* \right ) \S \Sd + g \left ( - \tfrac{1}{4} + {\gamma_2}  + \tfrac{1}{2} \bar{\gamma}_3  \right ) \S^2 \\ 
\hspace{10ex}+g^* \left ( - \tfrac{9}{4} + {\gamma_3}  + \tfrac{1}{2} \bar{\gamma}_2 + 2 c^* \right ) \Sd^2 \bigg ]
+ c^* \D_{\a} \Sd \left ( \S - 2 g^* \Sd \right )\sn~,
\eea
where
\bea{ll}
{\gamma_1} =  \tfrac{ \left ( -5+8c^*-4c \right ) f^2 + \left ( 4+4 \lambda - 8 \lambda^* - 4 c^* \right ) f + 4 \lambda^* +1 }{ 2 \left ( 3 f - 1 \right ) \left ( f - 1 \right ) }~, \\
 {\gamma_3} =  \tfrac{ \left ( 17 - 16 c^*   \right ) f^2 + \left ( 16 \lambda^* + 8 c^* - 8 \right ) f - 8 \lambda^*  - 1}{ 2 \left ( 3 f - 1 \right )  \left (  f - 1 \right ) }~,\\
{\gamma_2} =  \tfrac{\left ( 8 c - 7 \right ) f^2 - 8 \lambda  f -1}{ 2 \left ( 3 f - 1 \right ) \left ( f - 1 \right ) }~.
\eea
\item \underline{Conformal case}: For this case, equation (\ref{dgXc}) is not satisfied unless $g=0,~\lambda=-2/3$ and the supercurrent is
\bea{ll}
\N_{\a \ad} =~ i \pa_{\a \ad} \Sd \S  + \D_{\a} \Dd_{\ad} \left ( \tfrac{1}{3} \Sd \S \right )  + c.c. ~. \n  
\eea
\end{enumerate}
As in the case of old-minimal, it is straightforward to verify that all the above-mentioned supercurrents satisfy their respective conservation equations as they were presented in the previous section.

B.~\underline{\textbf{Higher derivative interacting theory}}:\\
For our second example, we consider a system that introduces interactions through
higher derivative terms. The Lagrangian we examine in this case is:
\bea{l}\n\label{ex2}
\mathcal{L}_{o}=-\S\Sd+g\D^{\a}\S~\D_{\a}\S+g^*\Dd^{\ad}\Sd~\Dd_{\ad}\Sd~.
\eea
The first thing we must do is to find the $X$ superfield. This is not as straightforward as in the previous case and we have to employ some superspace algebra, like the following identity
\bea{ll}\n
i\pa_{\a \ad} \S \D^{\b} Y_{\b} - c.c. &=~ \tfrac{1}{2} \D^{\b} \left\{ i\pa_{(\a \ad} \S Y_{\b)} - \tfrac{1}{3} \D_{(\a}\left [\Dd_{\ad} \S Y_{\b)} \right ] \right \} - c.c.\\ 
&+~ \tfrac{1}{2} \D_{\a} \left [ \Dd_{\ad} \S \D^{\b} Y_{\b} \right ] +
\tfrac{1}{2} \Dd_{\ad} \left [\D_{\a} \S \D^{\b} Y_{\b} \right ] - c.c. \\
&-~ \tfrac{1}{4} \D_{\a}\Dd_{\ad} L -c.c. ~,
\eea
where $Y_{\a} = \D_{\a} \S, ~ L = \D^{\a} \S ~ \D_{\a} \S $. With this in mind, we get that
\bea{l}\n
X =  -\tfrac{1}{2}\S \bar \S +\tfrac{g}{2} L~,~\Omega_{\b\a\ad}=-ig\pa_{(\b\ad}\S ~Y_{\a)}+\tfrac{g}{3}\D_{(\b}\left[\Dd_{\ad}\S ~Y_{\a)}\right]~. 
\eea
So, now we go through the list of the various supergravities and check whether this theory can be coupled to them and what are the supercurrents:
\begin{enumerate}
\item \underline{Old-minimal}: 
Equation (\ref{omXc}) is always satisfied, thus this theory can be coupled to
old-minimal supergravity. The corresponding supercurrents are
\bea{ll}\n
\hspace{-2ex}\N_{\a \ad} &=~  \D_{\a} \Dd_{\ad} \left \{ \tfrac{1}{3} \Sd \S  - \tfrac{5}{6} g L + \tfrac{5}{3} g^* \bar L + \tfrac{4}{3} \left ( 2 g \S \D^{\b} Y_{\b} - g^* \Sd \Dd^{\bd} \bar Y_{\bd} \right )\right \}  + c.c. \nonumber\\
&\hspace{2ex}+~ i \pa_{\a \ad} \Sd \left ( \S  + 2  g^*  \Dd^{\bd} \bar Y_{\bd} \right )  + g^* \Dd_{\ad} \left \{ \D_{\a} \Sd \Dd^{\bd} \bar Y_{\bd} \right \}  + c.c.\sn\\ 
&\hspace{2ex}+~ g \D^{\b} \left\{- i\pa_{(\a \ad} \S Y_{\b)} + \tfrac{1}{3} \D_{(\a}\left (\Dd_{\ad} \S Y_{\b)} \right ) \right \} + c.c.~,\\
\hspace{-2ex}\mathscr{M} &=-\left ( \tfrac{2}{3} + \lambda \right ) \S \Sd - 2g\left ( \tfrac{2}{3}   +  \lambda \right ) g \S \D^{\a} Y_{\a}  + \tfrac{2}{3} g^* \Sd  \Dd^{\ad} \bar Y_{\ad}\sn\\
&\hspace{2ex} + \tfrac{2}{3} g L - \tfrac{1}{3} g^* \bar L~.
\eea
\item \underline{New-minimal}: The coupling to new-minimal (\ref{newmXc}) is more restrictive, but it can still be done if we select $\lambda=-3/2,~Z_{\ad}=-g^*\Sd~\Dd_{\ad}\Sd$. The supercurrents for this case are
\bea{ll}\n
\N_{\a \ad} &=~ \D_{\a} \Dd_{\ad} \left \{ 2 \Sd \S  + \tfrac{1}{2} g L + 3 g^* \bar L + 6  g \S \D^{\b} Y_{\b} + 2 g^* \Sd \Dd^{\bd} \bar Y_{\bd} \right \}  + c.c. \nonumber\\
&\hspace{2ex}+~ i \pa_{\a \ad} \Sd \left ( \S  + 2  g^*  \Dd^{\bd} \bar Y_{\bd} \right )  + g^* \Dd_{\ad} \left \{ \D_{\a} \Sd \Dd^{\bd} \bar Y_{\bd} \right \}  + c.c.\sn\\
&\hspace{2ex}+~ g \D^{\b} \left\{- i\pa_{(\a \ad} \S Y_{\b)} + \tfrac{1}{3} \D_{(\a}\left (\Dd_{\ad} \S Y_{\b)} \right ) \right \} + c.c.~,\\
\mathscr{R} &=~  -\tfrac{5}{2} \S \Sd - 5 \left ( g \S \D^{\a} Y_{\a}  +  g^* \Sd  \Dd^{\ad} \bar Y_{\ad} \right ) - 2 g L - 2 g^* \bar L\sn~.
\eea
Notice that the fixed value $\lambda = -3/2$ corresponds to a zero R-charge for $\D_\a\S$ which is the only value for which the action is R-symmetry invariant.
\item \underline{New-new-minimal}: Coupling to new-new-minimal is not possible for this case, since (\ref{newnewmXc}) is too restrictive and has no solution.
\item \underline{Non-minimal}: For non-minimal, equation (\ref{nonmXc}) is always satisfied, therefore we can couple the theory to non-minimal supergravity. The supercurrents are
{\allowdisplaybreaks
\bea{l}
\n
\N_{\a \ad}=~ i \pa_{\a \ad} \Sd \left ( \S  + 2  g^*  \Dd^{\bd} \bar Y_{\bd} \right )  + g^* \Dd_{\ad} \left \{ \D_{\a} \Sd \Dd^{\bd} \bar Y_{\bd} \right \}  + c.c.\sn\\
\hspace{7ex} +\Dd_{\ad} \D_{\a} \Big [ \left ( \tfrac{1}{2} + {\gamma_1} \right )  \Sd \S  + \left ( {\gamma_4} - 1 \right ) g L + {\gamma_5} g^* \bar L\\
\hspace{17ex}+{\gamma_2}  g \S \D^{\b} Y_{\b} + {\gamma_3} g^* \Sd \Dd^{\bd} \bar Y_{\bd} \Big ]+c.c.\\
\hspace{7ex}+g \D^{\b} \left\{- i\pa_{(\a \ad} \S Y_{\b)} + {\tfrac{1}{3}} \D_{(\a}\left (\Dd_{\ad} \S Y_{\b)} \right ) \right \} + c.c.~,\\
\mS_{\a}=~\D_{\a} \bigg [ \left ( \tfrac{5}{4}  + {\gamma_1} + \tfrac{1}{2} \bar {\gamma}_1  - c^* \right ) \S \Sd + g \left ( {\gamma_2}  + \tfrac{1}{2} \bar{\gamma}_3  \right ) \S \D^{\b} Y_{\b} \sn\\
\hspace{10ex}+g^* \left ( {\gamma_3}  + \tfrac{1}{2} \bar{\gamma}_2 + 2 -2c^* \right ) \Sd \Dd^{\ad} \bar Y_{\ad} + \left (  \tfrac{1}{4} + {\gamma_4} + \tfrac{1}{2} \bar {\gamma}_5 \right ) g L\\
\hspace{10ex}+\left ( - \tfrac{1}{2} + {\gamma_5} + \tfrac{1}{2} \bar {\gamma}_4 \right ) g^* \bar L \bigg ]\\
\hspace{6ex}+ c^* \D_{\a} \Sd \left ( \S + 2 g \Dd^{\ad} \bar Y_{\ad} \right )~,
\eea
}
where the constants $\gamma$ are
\bea{ll}
{\gamma_1} =  \tfrac{ \left ( -5+8c^*-4c \right ) f^2 + \left ( 4+4 \lambda - 8 \lambda^* - 4 c^* \right ) f + 4 \lambda^* +1}{ 2 \left ( 3 f - 1 \right ) \left ( f - 1 \right ) }, \\
~  {\gamma_2} =  \tfrac{ 4 f \left ( \lambda +f -c f \right ) }{ \left ( 3 f - 1 \right ) \left ( f - 1 \right ) },
~ {\gamma_3} = - \tfrac{ 4 \left ( 2 f - 1 \right )  \left ( \lambda^* +f - c^* f  \right ) }{ \left ( 3 f - 1 \right )  \left (  f - 1 \right ) },\\
~  {\gamma_4} =  - \tfrac{ 4 f^2}{ 2 \left ( 3 f - 1 \right ) \left ( f - 1 \right ) },
~ {\gamma_5} =  \tfrac{  5 f^2 -1}{ 2 \left ( 3 f - 1 \right )  \left (  f - 1 \right ) }~.
\eea
\item \underline{Conformal case}: It is easy to check from (\ref{dgXc}) that the coupling of this theory (with nonzero $g$) to conformal supergravity is not possible.
\end{enumerate}
%

C.~\underline{\textbf{Interacting theory with spontaneous SUSY breaking}}:\\
Our third and last example is an interacting theory with a higher derivative term containing four supercovariant derivatives
\bea{l}\n\label{ex3}
\mathcal{L}_{o}=-\S\Sd+g\D^{\a}\S~\D_{\a}\S~\Dd^{\ad}\Sd~\Dd_{\ad}\Sd~.
\eea
This model has been studied in \cite{susybr5, susybr3, susybr4} where it has been demonstrated that the equations of motions have several vacuum solutions including one which breaks supersymmetry. For that reason it would be interesting to find the supercurrents for this theory and the list of supergravities it can be coupled to.

Similarly with the previous example, in order to find $X$ we have to use the following identity
\bea{ll}\n
i\pa_{\a \ad} \S \D^{\b} Y_{\b} - c.c. &=~ \tfrac{1}{2} \D^{\b} \left\{ i\pa_{(\a \ad} \S Y_{\b)} - \tfrac{1}{3} \D_{(\a}\left [\Dd_{\ad} \S Y_{\b)} \right ] \right \} - c.c.\\ 
&+~ \tfrac{1}{2} \D_{\a} \left [ \Dd_{\ad} \S \D^{\b} Y_{\b} \right ] +
\tfrac{1}{2} \Dd_{\ad} \left [\D_{\a} \S \D^{\b} Y_{\b} \right ] - c.c. ~,
\eea
where $Y_{\a} = \D_{\a}\S~\Dd^{\ad}\Sd~\Dd_{\ad}\Sd, ~ L = \D^{\a}\S~\D_{\a}\S~\Dd^{\ad}\Sd~\Dd_{\ad}\Sd $. The result is
\bea{l}\n
X = - \tfrac{1}{2} \S \bar \S~,~\Omega_{\b\a\ad}=-ig\pa_{(\b\ad}\S ~Y_{\a)}+\tfrac{g}{3}\D_{(\b}\left[\Dd_{\ad}\S ~Y_{\a)}\right]~.
\eea
This is perhaps surprising since the answer for $X$ does not depend on the coupling constant $g$ and is the same as the free theory.
\begin{enumerate}
\item \underline{Old-minimal}: Checking equation (\ref{omXc}), we conclude that we can couple the theory to old-minimal supergravity and the calculation of the supercurrents is straight forward. We get:
\bea{l}\n
\N_{\a \ad}=~  i \pa_{\a \ad} \Sd \left ( \S  + 2  g  \Dd^{\bd} \bar Y_{\bd} \right ) + \tfrac{1}{3} \D_{\a} \Dd_{\ad} \left ( \Sd \S + 8 g  \S \D^{\b} Y_{\b} - 4 g \Sd  \Dd^{\bd} \bar Y_{\bd} \right ) + c.c. \nonumber\\
\hspace{6ex}+g \Dd_{\ad} \left \{\D_{\a} \S \D^{\b} Y_{\b} + \D_{\a} \Sd \Dd^{\bd} \bar Y_{\bd} \right \}  + c.c.\sn\\ 
\hspace{6ex}+ g \D^{\b} \left\{- i\pa_{(\a \ad} \S Y_{\b)} + \tfrac{1}{3} \D_{(\a}\left (\Dd_{\ad} \S Y_{\b)} \right ) \right \} + c.c.~,\\
\mathscr{M}=~ - \left ( \tfrac{2}{3} + \lambda \right ) \S \Sd -2g\left(\tfrac{2}{3}+\lambda\right)\S \D^{\a} Y_{\a} +\tfrac{2}{3}g \Sd  \Dd^{\ad} \bar Y_{\ad}\sn ~.
\eea
\item \underline{New-minimal}: Coupling to new-minimal is possible only for $\lambda=-1$ and the supercurrents are
\bea{l}\n
\N_{\a \ad}=~  i \pa_{\a \ad} \Sd \left ( \S  + 2  g  \Dd^{\bd} \bar Y_{\bd} \right ) + \D_{\a}\Dd_{\ad} \left (\Sd \S + 4 g \S \D^{\b}  Y_{\b} \right )  + c.c.\nonumber\\
\hspace{7ex}+~  g \Dd_{\ad} \left \{\D_{\a} \S \D^{\b} Y_{\b} + \D_{\a} \Sd \Dd^{\bd} \bar Y_{\bd} \right \}  + c.c.\sn\\ 
\hspace{7ex}+~ g \D^{\b} \left\{- i\pa_{(\a \ad} \S Y_{\b)} + \tfrac{1}{3} \D_{(\a}\left (\Dd_{\ad} \S Y_{\b)} \right ) \right \} + c.c.~,\\
\mathscr{R}=-\S \Sd - 2 g \left( \Sd \Dd^{\ad} \bar Y_{\ad} +  \S \D^{\a} Y_{\a} \right ) \sn~.
\eea
\item \underline{New-new-minimal}: The coupling to new-new-minimal, as can be checked by (\ref{newnewmXc}) is not possible.
\item \underline{Non-minimal}: Coupling to non-minimal is possible and the supercurrents are
\bea{l}\n\label{ex3-supc}
\N_{\a \ad}=~  i \pa_{\a \ad} \Sd \left ( \S  + 2  g  \Dd^{\bd} \bar Y_{\bd} \right )+c.c.\sn\\
\hspace{6ex}+\Dd_{\ad} \D_{\a} \left \{ \left ( \tfrac{1}{2} + {\gamma_1} \right ) \Sd \S + g {\gamma_2} \S \D^{\b} Y_{\b} +  g{\gamma_3} \Sd  \Dd^{\bd} \bar Y_{\bd} \right \}  + c.c.\\
\hspace{6ex}+ g \Dd_{\ad} \left \{\D_{\a} \S \D^{\b} Y_{\b} + \D_{\a} \Sd \Dd^{\bd} \bar Y_{\bd} \right \}  + c.c.\\ 
\hspace{6ex}+ g \D^{\b} \left\{- i\pa_{(\a \ad} \S Y_{\b)} + \tfrac{1}{3} \D_{(\a}\left (\Dd_{\ad} \S Y_{\b)} \right ) \right \} + c.c.~,\\
\mS_{\a}=~\D_{\a} \bigg [ \left ( \tfrac{5}{4}  + {\gamma_1} + \tfrac{1}{2} \bar {\gamma}_1  - c^* \right ) \S \Sd + g \left ( {\gamma_2}  + \tfrac{1}{2} \bar{\gamma}_3  \right ) \S \D^{\b} Y_{\b}\sn\label{ex3-nm-S}\\
\hspace{11ex} +g \left (2+ {\gamma_3}  + \tfrac{1}{2} \bar{\gamma}_2 - 2 c^* \right ) \Sd \Dd^{\ad} \bar Y_{\ad} \bigg ]
+ c^* \D_{\a} \Sd \left ( \S + 2 g \Dd^{\ad} \bar Y_{\ad} \right )~,
\eea
with the $\gamma$ constants defined as:
\bea{ll}
{\gamma_1} =  \tfrac{ \left ( -5+8c^*-4c \right ) f^2 + \left ( 4+4 \lambda - 8 \lambda^* - 4 c^* \right ) f + 4 \lambda^* + 1}{ 2 \left ( 3 f - 1 \right ) \left ( f - 1 \right ) }, \\
~  {\gamma_2} =  \tfrac{ 4 f \left ( \lambda +f - c f \right ) }{ \left ( 3 f - 1 \right ) \left ( f - 1 \right ) },
~ {\gamma_3} = - \tfrac{ 4 \left ( 2 f - 1 \right )  \left ( \lambda^* +f - c^* f  \right ) }{ \left ( 3 f - 1 \right )  \left (  f - 1 \right ) }.
\eea
\item \underline{Conformal case}: The last case on the list, is this degenerate coupling to supergravity. Due to (\ref{dgXc}) we conclude that this is not possible.
\end{enumerate}

As we mentioned previously, the theory (\ref{ex3}) has a solution that spontaneously breaks supersymmetry \cite{susybr3,susybr4}. The reason why this can happen is the non-minimal nature of the complex linear supermultiplet. It has a bigger set of non-dynamical auxiliary components ($F\sim\D^2\S|,~P_{\a\ad}\sim \Dd_{\ad}\D_{\a}\S|,~\lambda_{\a}\sim \D_{\a}\S|,~\chi_{\a}\sim \Dd^{\gd}\D_{\a}\Dd_{\gd}\Sd|$) which for the free theory they must vanish on-shell. However, some special type of higher derivative theories can generate potential energy terms for some of these auxiliary fields without any kinetic energy terms. As a consequence, the on-shell equations of motion for these components remain algebraic but now they can have non zero solutions. These non-zero solutions force the auxiliary fields to acquire a non-zero v.e.v. and as a result, they break supersymmetry spontaneously. This mechanism happening or not depends a lot on the precise component structure of the action. For instance, if we interchange $\S$ and $\Sd$ in (\ref{ex3}) we get
\bea{l}\n\label{ex4}
\mathcal{L}_{o}=-\S\Sd+g\D^{\a}\Sd~\D_{\a}\Sd~\Dd^{\ad}\S~\Dd_{\ad}\S~.
\eea
which does not break supersymmetry. It is therefore of interest to see how this seemingly small change affects the supercurrent of the theory. Of course, the difference arises from the different component structure of the two theories, but we would like to find indications of this without having to do the detailed component projection.
 
When a spontaneous supersymmetry breaking solution exists, the fermionic version of the Goldstone theorem applies and we are expecting the existence of a massless fermion, a goldstino. An easy way to identify the goldstino is to look at the set of previously auxiliary fermionic degrees of freedom and find the combination that has
an algebraic term of the v.e.v acquiring bosonic component in his supersymmetry transformation ($
\delta_{S}\lambda_{\a}\sim \epsilon_{\a}\bar{F}+\dots
$). This is known as a \emph{shift} term. Of course this transformation has to be generated by the conserved charge of the supercurrent. Therefore, we should be able to identify the structure responsible for the supersymmetry breaking inside the supercurrent. Based on equation (\ref{cfc}) we conclude that the supercharges will have terms proportional to $\mS^{(1,2)}_{\a}$. From example C and equation (\ref{ex3-nm-S}) we find that the superfield $\mS$ has the structure
\bea{l}
\mS=A_1~\S\Sd~+gA_2~\S\D^{\b}Y_{\b}~+gA_3~\Sd\Dd^{\bd}\bar{Y}_{\bd}~,\n
\eea  
for some coefficients $A_i$ and coupling constant $g$. Therefore the component $\mS^{(1,2)}_{\a}=-\tfrac{1}{2}\left\{\Dd^2,\D_{\a}\right\}\mS|$ will include, among other terms, the following
\bea{l}
\mS^{(1,2)}_{\a}=-A_1\lambda_{\a}\bar{F}+4g(A_2+A_3)\lambda_{\a}F\bar{F}^2+\dots ~,
\eea
which through the Poisson bracket would generate the transformation with the shift term. Repeating the same calculation for the non supersymmetry breaking theory (\ref{ex4}), we can easily show that due to the linearity constraint of $\S$ all the corresponding terms proportional to $g$ vanish. This illustrates the difference between the two theories and the potential of the first one to break supersymmetry.

Moreover, when the supersymmetry breaking theory is coupled to a supergravity the goldstino will be eaten by the gravitino in order for the gravitino to become massive. For the theory in example C, the origin of the goldstino can be traced back to the auxiliary fermions of $\S$. However if the supergravity we couple the theory to, is non-minimal (it also has auxiliary fermions) and we allow higher derivative terms in the supergravity sector, then there is a possibility that the goldstino mode will be provided by a combination of the previously auxiliary fermions of $\S$ and the auxiliary fermions of the non-minimal supergravity. It would be interesting to see explicitly this mechanism at the component level of the theory. Specifically, we would like to see the coupling to the gravitino. Of course, since we are working in the linear approximation we will not be able to see the gravitino mass term emerging but at least we can see which components of the supercurrents will play a role. Following \cite{NM1, NM2, NM3}\footnote{For ease of calculation we go to a Wess-Zumino gauge and select $f=0$.}
we have the following definitions for the fermionic fields of non-minimal supergravity
\bea{l}\n
\tfrac{1}{2!}\Dd^2\D_{(\r}H_{\a)\ad}|=\tfrac{1}{\sqrt{2}}\psi_{\r\a\ad}~,~
\D^2\chi_{\a}|=-\tfrac{1}{2\sqrt{2}}\psi_{\a}~,\\
\Dd_{\ad}\D^{\a}\chi_{\a}|=\tfrac{1}{2}\bar{\r}_{\ad}-\tfrac{1}{\sqrt{2}}\bar{\psi}_{\ad}~,~
\Dd^2\D^2\chi_{\a}|=\b_{\a}+i\pa_{\a}{}^{\ad}\left(\tfrac{1}{4}\bar{\r}_{\ad}-\tfrac{1}{\sqrt{2}}\bar{\psi}_{\ad}\right)~.
\eea
Therefore the fermionic part of the interaction term (\ref{int-nonm}) is
\bea{ll}\n\label{comp-int-nonm}
S_{int}|_{F}=\int d^4x ~& \tfrac{1}{2\sqrt{2}M}\psi^{\b\a\ad}\[\tfrac{1}{2!}\D_{(\b}\N_{\a)\ad}|\]+c.c.\\
&-\tfrac{1}{2\sqrt{2}M}\psi^{\a}\[\Dd^2\mS_{\a}|+2i\pa_{\a}{}^{\ad}\bar{\mS}_{\ad}|\]+c.c.\\
&+\tfrac{1}{M}\b^{\a}\[\mS_{\a}|\]+\tfrac{1}{4M}\r^{\a}\[\Dd^{\ad}\D_{\a}\bar{\mS}_{\ad}|\]+c.c.~.
\eea
%
%
%
\section{Conclusion}
~~~~The complex linear supermultiplet is a well known variant representation of the scalar supermultiplet. However since it describes the same physical degrees of freedom as the simpler chiral multiplet, it is not studied as extensively. Nevertheless, there are situations where a theory defined in terms of complex linear superfields cannot be equally described in terms of chiral superfields. A prime example is the spontaneous supersymmetry breaking generated by higher derivative terms.

Motivated by the above, we investigated the supercurrent multiplet of a generic $4D,~\mathcal{N}=1$ theory of complex linear superfield and aspects of its coupling to supergravity. Using the Noether procedure we find
explicit expressions for the supercurrents of the arbitrary $\S$ theory and verify that they satisfy the appropriate superspace conservation equations, which we have derived from the on-shell limit of the Bianchi identities. We also presented the component projection of these conservation equations to spacetime. These spacetime conservation equations define the energy-momentum tensor, its fermionic superpartner and the $U(1)_R$ current for the generic theory of complex linear superfield.

Furthermore, rigid Super-Poincar\'{e} invariance gives rise to a special superfield $X$. We find that the role of this object is to decide whether the coupling of the theory to a specific formulation of supergravity is possible or not. The algorithm to do that is very simple: step 1. For any given theory, calculate the superfield $X$ based on (\ref{K}), step 2. Check if $X$ as calculated in step 1 satisfies any of the four conditions (\ref{omXc}), (\ref{newmXc}), (\ref{newnewmXc}), (\ref{nonmXc}). If it does, then the theory can be coupled to the corresponding supergravity formulation. Due to the special type of constraints (\ref{omXc}) and (\ref{nonmXc}), the result is that any theory can be coupled to old-minimal and non-minimal supergravity\footnote{For example, we can appropriately choose $\lambda$ so these constraints are satisfied}. On the other hand, not every theory can be coupled to new-minimal and new-new-minimal formulation of supergravity as we have seen in section 7. Although the existence of superfield $X$ was known already, its crucial role for coupling to supergravities, has not been recognised until now. Our proposed method is based upon this observation and as far as we know, there is no other known process (apart from trial and error) in order to deduce which formulation of supergravity is consistent with a given theory.

In addition, we have illustrated how the conditions that permit coupling to new-minimal supergravity are exactly the ones that permit the realization of the $U(1)_R$ symmetry. The connection of new-minimal supergravity and R-symmetry is known, due to the fact that the conservation equation of new-minimal supergravity results in the spacetime conservation of the vector supercurrent. However, the new contribution here is the explanation and deeper understanding of this connection. It comes from comparing the Noether procedure for coupling to new-minimal supergravity with the one for R-symmetry invariance. Also due to this connection, one can use our $X$-method and equation (\ref{newmXc}) to straightforwardly determine whether a theory has R-symmetry or not.

Finally, we apply the above results to specific examples of theories. Among these examples two of them were defined including higher derivative terms of similar kind but one of them has supersymmetry breaking solutions. This was illustrated by calculating part of the supersymmetry charge in terms of components. Also for the supersymmetry breaking case we calculated the component interaction for the fermion of the $\S$ theory with the fermions of  non-minimal supergravity.

We hope, that these results will contribute to the better understanding of other systems where complex linear superfields are being used.

\section{Acknowledgments}
This work was supported by the grant P201/12/G028 of the Grant agency of Czech Republic. We are grateful for discussions with Ulf Lindstr\"om and Martin Ro\v{c}ek.
%

%
%
\footnotesize{

}

\end{document}